\documentclass[11pt]{article}
\usepackage{graphics}
\usepackage[dvips]{graphicx}
\usepackage{mathrsfs}
\usepackage{amssymb}
\usepackage{verbatim}
\usepackage{float}
\newcommand{\be}{\begin{equation}}
\newcommand{\ee}{\end{equation}}
\newcommand{\bea}{\begin{eqnarray}}
\newcommand{\eea}{\end{eqnarray}}
\newcommand{\nnmb}{\nonumber}
\newcommand{\NPB}[3]{\emph{ Nucl.~Phys.} \textbf{B#1} (#2) #3}   
\newcommand{\PLB}[3]{\emph{ Phys.~Lett.} \textbf{B#1} (#2) #3}   
\newcommand{\PRD}[3]{\emph{ Phys.~Rev.} \textbf{D#1} (#2) #3}

\newcommand{\HPA}[3]{\emph{ Helv.~Phys.~Acta} \textbf{#1} (#2) #3}

\def\4vol{{\int d^4x \sqrt{-g}}}
\def\lag{{\mathscr{L}}}

\def\del{{\partial}}

\def\xs{{\varphi_s}}
\def\xd{{\varphi_1}}
\def\xu{{\varphi_2}}
\def\xm{{\varphi}}

\def\simgt{\stackrel{>}{{}_\sim}}
\newcommand{\nc}{\newcommand}

\nc{\ra}{\rightarrow}
\nc{\lsim}{\begin{array}{c}\,\sim\vspace{-21pt}\\< \end{array}}
\nc{\gsim}{\begin{array}{c}\sim\vspace{-21pt}\\> \end{array}}
\nc{\nt}{\tilde{N}}

\nc{\LL}{L}
\nc{\vv}{\tilde{v}}

\title{
\vspace*{-1.3cm}
\begin{flushright}
\normalsize{
ANL-HEP-PR-04-38\\
EFI-04-09\\
  }
\end{flushright}
\vspace{1.5cm}
\Large
\textbf{Electroweak Baryogenesis and Dark Matter in the nMSSM}
\vspace*{1.0cm}
\author{\large\textbf{A.~Menon$^{a,b}$}, \textbf{D.E.~Morrissey
$^{a,b}$},
and \textbf{C.E.M.~Wagner$^{a,b}$}\\ \\[0.5cm]
$^a$\normalsize\emph{HEP Division, Argonne National Laboratory,
9700 Cass Ave.,
Argonne, IL 60439, USA} \\
$^b$\normalsize\emph{Enrico Fermi Institute, Univ. of Chicago,
5640 S. Ellis Ave., Chicago, IL 60637, USA}}} 
\begin{document}
\setcounter{page}{0}
\maketitle
\vspace*{1cm}
\begin{abstract}

We examine the possibility of electroweak baryogenesis and dark
matter in the nMSSM, a minimal extension of the MSSM 
with a singlet field.
This extension avoids the usual domain wall problem of the NMSSM,
and also appears as the low energy theory in models of dynamical
electroweak symmetry breaking with a so-called fat-Higgs boson.
We demonstrate that a strong, first order electroweak phase transition,
necessary for electroweak baryogenesis,
may arise in regions of parameter space where the lightest neutralino
provides an acceptable dark matter candidate. We investigate the
parameter space in which these two properties are fulfilled and
discuss the resulting phenomenology.
In  particular, we show that there are always two light 
CP-even and one light CP-odd Higgs bosons with
masses smaller than about 250~GeV. Moreover,  
in order to obtain a realistic relic density,
the lightest neutralino mass tends to be 
smaller than $M_Z/2$, in which case the lightest Higgs boson
decays predominantly into neutralinos.  
\end{abstract}

\thispagestyle{empty}
\newpage

\setcounter{page}{1}

\section{Introduction}

The Standard Model (SM) provides an excellent description of all
elementary particle interactions up to energies of order 100~GeV.
However, there are several reasons to expect that an extension of
the Standard Model description is needed at energies 
slightly above this scale. The most important of these is 
the precise mechanism for the generation of elementary 
particle masses. In the Standard Model, the Higgs mechanism 
provides a self-consistent parametrization of this mechanism, 
but relies on mass parameters 
that are sensitive to ultraviolet scales at the quantum level. 
Two ways to avoid this problem are to invoke 
supersymmetry, or to break the electroweak symmetry dynamically.
In both cases, the new physics provides an effective energy 
cutoff for the SM description that is of order of the weak scale. 
While the construction of 
supersymmetric extensions consistent with
low energy observables is a relatively simple 
task~\cite{Haber:1984rc,Martin:1997ns},
this is much more difficult to do in models 
that break the electroweak
symmetry dynamically due to the non-perturbative 
nature of the interactions in the symmetry breaking 
sector~\cite{Hill:2002ap}.

The Standard Model description also fails to answer two of
the most important questions at the interface of
particle physics and cosmology: the nature of the dark matter
and the origin of the baryon asymmetry.  
A consistent dark matter density may  easily 
arise in the presence of stable, neutral, weakly
interacting particles with masses of order of the weak
scale. In this sense, the observation of a non-vanishing
dark matter density provides an additional motivation for
the existence of new physics at the weak scale. 
The lightest supersymmetric particle (LSP) in models of
low-energy supersymmetry with R-Parity conservation provides
a natural source for the observed dark-matter 
density~\cite{Haber:1984rc,Martin:1997ns}.

There are several ways of explaining the origin
of the matter-antimatter asymmetry. Some of them rely on physics
at scales much larger than the weak scale. 
Leptogenesis, for instance, provides a mechanism
for the generation of a lepton asymmetry via the decay of  
heavy Majorana neutrinos which explains the smallness of the neutrino
masses in the See-saw framework. This lepton asymmetry is 
subsequently transformed into a baryon asymmetry via anomalous
weak interaction processes: baryogenesis proceeds 
from leptogenesis~\cite{leptogen,BBP03}.
Due to the high-energy nature of the fundamental interactions leading
to leptogenesis, it is difficult to test the realization of
this scenario by weak-scale experiments.

Electroweak baryogenesis~\cite{EWBGreviews} provides an 
alternative to models of baryogenesis
at high-energies, relying only on physics at the weak scale. 
The realization of this scenario demands that the anomalous
baryon-number violating processes are out of equilibrium
in the broken phase at the critical temperature of the
electroweak symmetry breaking phase transition. This is possible
only if the phase transition
is strongly first order, or equivalently,
\begin{equation}
\frac{\xm(T_c)}{T_c} \simgt 1,
\end{equation}
where $\xm(T_c)$ is the Higgs vacuum expectation value 
in the broken phase at the critical temperature $T_c$. 
Such a strongly first order transition
cannot be achieved within the SM framework.

Supersymmetric theories introduce new physics at the weak 
scale which may lead to a sufficiently strong first 
order phase transition.
This can be achieved even within the
minimal supersymmetric extension of the SM (MSSM). However,
in the minimal case, this demands Higgs masses only slightly 
above the present experimental bounds and a light stop with 
a mass below that of the top-quark~\cite{PT}. Therefore, this scenario 
is highly constrained and may be soon tested by the 
Tevatron collider experiments~\cite{Csaba}. 

As we will explain in the next section, there are good reasons
to go beyond the minimal supersymmetric framework. In this work
we shall study a minimal extension of the MSSM which includes 
a singlet field having restricted interactions. 
In Section~\ref{min}, after motivating this extension, 
we will describe its most relevant properties. 
In Section~\ref{tz} we will define the model precisely, 
set out our notation, and study the particle spectrum 
at zero-temperature as well as the collider constraints.
Section~\ref{ewbg} consists of an investigation of the strength of the
electroweak phase transition within the nMSSM.  In Section~\ref{dm}
we will study some of the cosmological implications of the model.
In particular, we will investigate the constraints on the model
that arise if it is to provide a realistic dark matter density. 
Section~\ref{pheno} consists of a discussion of the resulting
phenomenology based on the results of the previous sections.
We reserve section~\ref{conc} for our conclusions.

\section{Minimal Extension of the 
MSSM with a Singlet Superfield\label{min}}
The minimal 
supersymmetric extension of the SM stabilizes
the gauge hierarchy, leads naturally 
to a consistent unification of the gauge couplings near 
the Planck scale, and provides a viable dark matter candidate
if the neutralino, a superpartner of the neutral Higgs and gauge
boson particles, is the LSP~\cite{Haber:1984rc,Martin:1997ns}.
Despite these successes, the model has some 
unattractive features as well,
among which is the $\mu$-problem.  Namely, 
the $\mu$ term, $\mu\hat{H_1}\cdot\hat{H_2}$, must
be included in the superpotential, with $|\mu|$ of order of
the weak scale, if the electroweak symmetry is to be broken.
While the $\mu$ parameter is stable under quantum corrections
as a result of supersymmetry, it is difficult 
(although possible~\cite{GM}) 
to explain why this dimensionful quantity
should be so much smaller than~$M_P$ or~$M_{GUT}$.  

  A simple solution is to replace~$\mu$ 
by the VEV of a gauge singlet chiral superfield, $\hat{S}$. 
Other dimensionful couplings involving the singlet are 
then forbidden by demanding that the 
superpotential obey an additional symmetry.
In the most common formulation, the Next-to-Minimal 
Supersymmetric~SM~(NMSSM)~\cite{nmssm}, one imposes 
a~$\mathbb{Z}_3$ symmetry
under which the fields transform as 
$\hat{\Phi}_i\to\exp(2\pi iq_i/3)\;\hat{\Phi}_i$,
where the charges $q_i$ are given in table~\ref{charges}.
The most general renormalizable superpotential is then
\bea
 W_{ren} &=&  \lambda\,\hat{S}\,\hat{H}_1\cdot \hat{H}_2
     \;+\; \kappa\,\hat{S}^3\nonumber\\
     && \;+\; y_u\,\hat{Q}\cdot\hat{H}_2\;\hat{U}^c  
     \;+\; y_d\,\hat{Q}\cdot\hat{H}_1\;\hat{D}^c \;+\;
     y_e\,\hat{L}\cdot\hat{H}_1\;\hat{E}^c,
\label{wren}
\eea
where $\hat{S}$ is the singlet superfield and the other fields
are the same as in the MSSM.  Except for the cubic singlet 
self-coupling, this is just the MSSM superpotential with a field
dependent~$\mu$-term proportional to the singlet field.  
Without this additional cubic term, 
the superpotential is invariant under an anomalous U(1)$_{PQ}$, 
whose charges are listed in table~\ref{charges},
that gives rise to an unacceptable axion~\cite{Abel:1995uu}.  
The cubic term explicitly breaks U(1)$_{PQ}$ down to its maximal
$\mathbb{Z}_3$ subgroup, thereby removing the axion while still
forbidding all dimensionful couplings.  
Unfortunately, this generates new difficulties.
When the singlet acquires a VEV, necessarily near the 
electroweak scale,
the~$\mathbb{Z}_3$ symmetry is broken as well
producing cosmologically unacceptable domain 
walls~\cite{Abel:1995wk}.  
The domain wall problem can be avoided if $\mathbb{Z}_3$
violating non-renormalizable operators are included.  
However, such operators generate a large singlet tadpole 
that destabilizes the hierarchy~\cite{Bagger:1993ji}.

\begin{table}[hbt]
\begin{tabular}{|c|c|c|c|c|c|c|c|c|c|c|}
\hline
 &$\hat{H}_1$&$\hat{H}_2$&$\hat{S}$&$\hat{Q}$&$\hat{L}$&
$\hat{U}^c$&$\hat{D}^c$&$\hat{E}^c$&W\\
\hline
U(1)$_Y$&-1/2&1/2&0&1/6&-1/2&-2/3&1/3&1&0\\
\hline
$\mathbb{Z}_3\subset$U(1)$_{PQ}$&1&1&-2&-1&-1&0&0&0&0\\
\hline
U(1)$_R$&0&0&2&1&1&1&1&1&2\\
\hline
$\mathbb{Z}_5^R,\mathbb{Z}_7^R\subset$U(1)$_{R'}$&1&1&4&2&2&3&3&3&6\\
\hline
\end{tabular}
\caption{Charges of fields under the Abelian symmetries discussed
in the text.
\label{charges}}
\end{table}

  As shown in~\cite{Panagiotakopoulos:1999ah,Panagiotakopoulos:2000wp,
Dedes:2000jp}, both problems can be avoided in the context of 
an N=1 supergravity scenario.  In the absence of the cubic 
singlet term, the superpotential
of Eq.~(\ref{wren}) obeys the U(1)$_{PQ}$, and U(1)$_R$ symmetries 
listed in table~\ref{charges}, and so is also invariant under 
the group U(1)$_{R'}$ with charges ${R'} = 3{R} + {PQ}$.  
This symmetry alone is enough to give the superpotential 
of Eq.~(\ref{wren}) with no 
cubic term, as are the maximal $\mathbb{Z}^R_5$ and $\mathbb{Z}^R_7$ 
subgroups of U(1)$_{R'}$.  If we now demand that both the 
superpotential and the K\"ahler potential obey one of 
these discrete R-symmetries instead of the~$\mathbb{Z}_3$ 
symmetry of the NMSSM, the superpotential will be of the
desired form up to a possible singlet tadpole term.
Using power counting arguments it may be shown that a
singlet tadpole does arise, but only at the six~($\mathbb{Z}^R_5$) 
or seven~($\mathbb{Z}^R_5$) loop level. The loop suppression 
in both cases is large enough that the 
induced tadpole does not destabilize 
the hierarchy~\cite{Panagiotakopoulos:1999ah,
Panagiotakopoulos:2000wp}.  Therefore, this mechanism very
elegantly solves three problems: it prevents the appearance of 
dimensionful couplings (other than the tadpole) in the 
renormalizable part of the superpotential; the induced singlet
tadpole explicitly breaks U(1)$_{PQ}$ and its discrete subgroups, 
thereby avoiding unacceptable axions and domain walls; 
and the loop suppression of the tadpole 
leads naturally to a singlet VEV of the order of $M_{SUSY}$.
Following~\cite{Dedes:2000jp}, we shall refer to this 
model as the nearly Minimal Supersymmetric 
Standard Model~(nMSSM).  

Another interesting feature of the model is that something like
R-Parity arises from the imposed R-symmetries. 
These symmetries forbid the appearance of all $d=4\;$ 
$B$- and $L$-violating operators as well as the dominant 
higher dimensional $B$-violating operators that contribute
to proton decay.  While proton stability is ensured, there are
non-renormalizable operators that make the LSP unstable.
As will be shown in Section~\ref{cn}, the LSP of the model
under study is nearly always the lightest neutralino.
In the $\mathbb{Z}_7^R$ symmetric case, this symmetry forbids all 
$d\leq6$ operators that could lead to the decay of such an LSP,
and na\"ive dimensional analysis shows that it has a lifetime
in excess of the age of the universe.  
The issue is a bit more delicate
in the $\mathbb{Z}_5^R$ case since the $L$-violating $d=5$ operator
$\hat{S}\hat{S}\hat{H_2}\hat{L}$ is allowed by the symmetry.
We find that the lifetime of the neutralino LSP induced by this
operator is greater than the age of the Universe provided the 
cutoff scale (by which non-renormalizable operators are suppressed) 
exceeds $\Lambda \gtrsim 3\times 10^{14}$~GeV.  The details of 
our estimate are presented in Appendix~\ref{appb}.
Therefore, the same symmetries that ensure a 
natural solution to the $\mu$-problem stabilize the LSP, and 
provide the means for a sufficiently large proton
lifetime. 

   The nMSSM is quite attractive for a number of phenomenological 
reasons as well.  In the MSSM, large stop masses, 
high $\tan\beta$, and some fine-tuning are needed 
to evade the LEP-II Higgs mass bounds~\cite{Okada:1990vk}.
These bounds are much easier to avoid in the nMSSM and 
the NMSSM since there is an additional tree-level
contribution to the CP-even Higgs masses. 
The same LEP-II bounds on the lightest
neutral Higgs also put severe constraints on the parameter
space consistent with Electroweak 
Baryogenesis~(EWBG) within the~MSSM~\cite{Quiros:2000wk}.  
On the other hand, EWBG does appear to be 
possible within the NMSSM~\cite{Pietroni:in,Davies:1996qn,
Huber:2000mg} and other singlet extensions~\cite{Kang:2004pp} 
due to additional terms in the tree-level
potential.  We find that the same holds true for the nMSSM.

Finally, we note that the superpotential and soft-breaking
terms of the nMSSM also arise
as the low-energy effective theory of minimal
supersymmetric models with dynamical electroweak symmetry
breaking, the so-called \lq\lq Fat-Higgs \rq\rq 
models~\cite{Harnik:2003rs}.
In these models, the value of the Higgs-singlet  
coupling $\lambda$ is not restricted by the requirement of
perturbative consistency up to the grand-unification scale.
Instead, the precise value of $\lambda$  
depends on the scale of dynamical symmetry breaking, 
$\lambda$ being larger for smaller values of this scale. 
In our work, we shall focus on the case where 
perturbative consistency holds up to high-energy scales, 
but  we will also comment on how our results are modified if
we ignore the perturbativity constraint on $\lambda$.

\section{The nMSSM at Zero-Temperature\label{tz}}
  Much of the analysis and notation in this section follows
that of~\cite{Panagiotakopoulos:2000wp}.  For simplicity, 
we will include only the Higgs, singlet,
and third generation quark/squark fields in the superpotential. 
The superpotential, including the loop-generated tadpole term, 
is then
\be
W_{nMSSM}  = \lambda \hat{S} \hat{H}_1\!\cdot\!\hat{H}_2 
 \:+\: \frac{m_{12}^2}{\lambda}\hat{S} \:+\:
     y_t\hat{Q}\!\cdot\!\hat{H}_2\,\hat{U}^c \:+\: \ldots
\ee
where $\hat{H}_1^t = \left({\hat{H}_1^0, \hat{H}_1^-}\right)$,
$\hat{H}_2^t = \left({\hat{H}_2^+, \hat{H}_2^0}\right)$
denote the two Higgs superfields, $\hat{S}$ is the singlet superfield,
and $A\!\cdot\!B = \epsilon^{ab}A_aB_b$ with $\epsilon^{12} = 1$.

  The tree-level potential is then $V_0 = V_F+V_D+V_{soft}$:
\bea
V_F &=& |\lambda H_1\!\cdot\!H_2 \:+\: 
 \frac{m_{12}^2}{\lambda}|^2
\:+\: \left|\lambda S H^0_1 \:+\: 
 y_t\tilde{t}_L\tilde{t}_R^*\right|^2 \nonumber\\
&&\:+\: |\lambda SH_1^- + y_t\tilde{b}_L\tilde{t}_R^*|^2
+ |\lambda S|^2H_2^{\dagger}H_2 \nonumber\\
&&\:+\: |y_t\tilde{t}_R^*|^2H_2^{\dagger}H_2 
\:+\: |y_t\tilde{Q}\!\cdot\!H_2|^2,\nnmb\\
&&\nnmb\\
V_D &=& \frac{\bar{g}^2}{8}(H_2^{\dagger}H_2-H_1^{\dagger}H_1)^2
 \:+\: \frac{g^2}{2}|H_1^{\dagger}H_2|^2\\
&&\nnmb\\
V_{soft} &=& m_1^2H_1^{\dagger}H_1 \;+\; m_2^2H_2^{\dagger}H_2
\;+\; m_s^2|S|^2 \nonumber\\
&&\;+\; (t_s\,S + h.c.) \;+\; (a_{\lambda}S\,H_1\!\cdot\!H_2 + h.c.)
\nonumber\\
&& \:+\: m_Q^2\tilde{Q}^{\dagger}\tilde{Q} + m_U^2|\tilde{t_R}|^2
\:+\: (a_t\,\tilde{Q}\!\cdot\!H_2\,\tilde{t}_R^* + h.c.).\nnmb
\eea
In writing $V_D$ we have defined $\bar{g} = \sqrt{g^2+g'^2}
=g/\cos\theta_W$. 

  The couplings $a_{\lambda},t_s,\lambda,m_{12}^2,y_t,a_{t}$
can all be complex, but not all their phases are physical.
By suitable redefinitions of $\hat{S},\hat{H}_1$, and $\hat{H}_2$,
the parameters $\lambda$ and $m_{12}^2$ can both be made 
real~\cite{Ellis:1988er}.  
To simplify the analysis and to avoid spontaneous CP violation, 
we shall assume that the 
soft-breaking parameters $a_{\lambda},t_s,a_t$ and
the Yukawa $y_t$ are also real.\footnote{This assumption is 
not completely \textit{ad hoc}.  Within a minimal supergravity 
scenario, the soft breaking parameters are proportional to the 
corresponding terms in the superpotential.}
Moreover, we may take $a_{\lambda}$ and
$t_s$ to be positive  provided we allow $\lambda$ and
$m_{12}^2$ to have either sign.
Real parameters are not sufficient to exclude  
spontaneous CP violation, however, so we must check this 
explicitly.  We must also verify that the potential 
does not generate a VEV for either of the charged Higgs fields.  

If none of the squark fields get VEV's, the tree-level Higgs 
potential becomes
\bea
V_0 &=& m_1^2H_1^{\dagger}H_1 \;+\; m_2^2H_2^{\dagger}H_2 
\;+\; m_s^2|S|^2
\;+\; \lambda^2|H_1\!\cdot\!H_2|^2\nonumber\\
&& \;+\; \lambda^2|S|^2(H_1^{\dagger}H_1 \;+\; H_2^{\dagger}H_2)
\;+\; \frac{\bar{g}^2}{8}(H_2^{\dagger}H_2 - H_1^{\dagger}H_1)^2
\;+\; \frac{g^2}{2}|H_1^{\dagger}H_2|^2
\nonumber\\
&& \;+\; t_s(S + h.c.) \;+\; a_{\lambda}(S\,H_1\!\cdot\!H_2 + h.c.)
\;+\; m_{12}^2(H_1\!\cdot\!H_2 + h.c.).
\label{vhiggs}
\eea 
We may choose an SU(2)xU(1) gauge such that 
$\left<H_1^-\right> = 0$ and $\left<H_1^0\right> \in 
\mathbb{R}^{\geq}$ at the minimum of the potential.
Taking the derivative of $V_0$ with respect to $H_2^+$ and
evaluating the result at the minimum, we find
\be
\left.\frac{\partial V_0}{\partial H_2^+}\right|_{H_i = v_i} =
{v_+}^*\left[m_2^2 + \lambda^2|v_s|^2 + \frac{\bar{g}^2}{4}
(|v_2|^2-v_1^2) + \frac{g^2}{2}v_1^2 + \frac{\bar{g}^2}{8}
|v_+|^2\right],
\label{hcharge}
\ee
where we have defined $\left<H_2^0\right> = v_2$, 
$\left<H_2^+\right> = v_+$, and $\left<S\right> = v_s$.  
It follows that $\left<H_2^+\right>$
vanishes at the minimum provided $m_2^2 + \lambda^2|v_s|^2 + 
\frac{\bar{g}^2}{4}
(|v_2|^2-v_1^2) + \frac{g^2}{2}v_1^2 > 0$.
  
If the charged Higgs VEV's vanish at the minimum,
the only part of the potential that depends on the phases of 
the Higgs fields are the last three terms in Eq.~(\ref{vhiggs}):
\be
V_{phase} = t_s\,(S + h.c.) + a_{\lambda}(S\,H_1^0\,H_2^0 + h.c.)
+ m_{12}^2(H_1^0\,H_2^0 + h.c.).
\ee
Recalling that $a_{\lambda}$ and $t_s$ are both real and positive,
the potential will have an absolute minimum with $\left<S\right> = v_s
\in \mathbb{R}^{\leq}$ and $\left<H_2^0\right> = v_2 \in 
\mathbb{R}^{\geq}$
provided $m_{12}^2 <0$.  While this condition is sufficient
to avoid spontaneous CP violation, the result
of~\cite{Romao:jy} indicates that it is not necessary.
We will focus on the $m_{12}^2<0$ case because
it simplifies the analysis, and as we shall see below, is preferred 
by the constraints on the scalar Higgs masses.  
However, we have also examined the $m_{12}^2 >0$ case,
and find that once we impose the experimental constraints described 
in the following section, the parameter space with $m_{12}^2 >0$ is
very restricted, and tends to be inconsistent with 
electroweak baryogenesis.

  With $m_{12}^2 <0$, the field VEV's are all real 
and have fixed sign: 
\be
\begin{array}{ccc}
\left<S\right>=v_s<0\,&\phantom{aa}\left<H^0_1\right>=v_1>0,
&\phantom{aa}\left<H^0_2\right> = v_2 >0.
\end{array}
\ee
We define the angle $\beta$ as in the MSSM:
\be
\begin{array}{cc}
v_1 = v\cos\beta,&\phantom{aaaaa}v_2 = v\sin\beta, 
\end{array}
\ee
with $v \simeq 174$~GeV.  
We also define $\mu = -\lambda v_s$, since this is the 
quantity that corresponds to the $\mu$ parameter in the MSSM.
Note that $\mu$ can have either sign, depending on the sign of
$\lambda$.  

   The minimization conditions for $H_1^0$,~$H_2^0$,~and~$S$ 
can be used to relate the scalar soft masses to 
the other parameters in terms of the VEV's.
These give
\bea
m_1^2 &=& -(m_{12}^2 + a_{\lambda}v_s)\frac{v_2}{v_1} 
- \frac{\bar{g}^2}{4}(v_1^2-v_2^2) - \lambda^2(v_2^2 + v_s^2)
- \frac{1}{2v_1}\left.\frac{\del \Delta V}{\del H_1^0}
\right|_{H_1^0=v_1},
\nonumber\\
m_2^2 &=& -(m_{12}^2 + a_{\lambda}v_s)\frac{v_1}{v_2} 
+ \frac{\bar{g}^2}{4}(v_1^2-v_2^2) - \lambda^2(v_1^2 + v_s^2)
- \frac{1}{2v_2}\left.\frac{\del \Delta V}{\del H_2^0}
\right|_{H_2^0=v_2},
\nonumber\\
m_s^2 &=& -a_{\lambda}\frac{v_1v_2}{v_s} -\frac{t_s}{v_s} 
- \lambda^2\,v^2 - \frac{1}{2v_s}\left.\frac{\del \Delta V}{\del S}
\right|_{S=v_s},
\label{mincond}
\eea
where $\Delta V$ consists of contributions to the effective 
potential beyond tree-level.  To one-loop order
\be
 \Delta V = \frac{1}{(4\pi)^2}\left[\sum_{b}g_b\,h(m_b^2)
  - \sum_{f}g_f\,h(m_f^2)\right],
\ee
where the first sum runs over all bosons, the second over 
all Weyl fermions, $g_i$ is the number of (on-shell)
degrees of freedom, $m_i$ is the 
field-dependent mass eigenvalue,
and the function $h(m^2)$ is given by 
(in the $\overline{DR}$ scheme)
\be
  h(m^2) = \frac{m^4}{4}\left[\ln\left(\frac{m^2}{Q^2}\right)
 - \frac{3}{2}\right].
\ee
The one-loop corrections are therefore given by
\be
\Delta m_i^2 =\left. -\frac{2}{64\pi^2}\left(
\sum_bg_bm_b^2
\frac{\del m_b^2}{\del{H_i}^2}
\left[\ln(\frac{m_b^2}{Q^2})-1\right] 
 - \sum_fg_fm_f^2
\frac{\del m_f^2}{\del{H_i}^2}
\left[\ln(\frac{m_f^2}{Q^2})-1\right]
\right)\right|_{H_i=v_i}. 
\ee

\subsection{Charginos and Neutralinos\label{cn}}

The chargino and neutralino sectors provide important
phenomenological constraints on the model.  
The fermion component of the singlet superfield, the singlino, 
leads to a fifth neutralino state.  Assuming the sfermions
to be heavy, with masses of order a few hundred GeV, 
and values of $\lambda$ that remain perturbative up to 
a grand unification scale of order $10^{16}$~GeV,
the LSP of the model is always the lightest neutralino 
with a mass below about 60~GeV.
 
  The chargino mass matrix, in the basis 
$(\tilde{W}^+,\tilde{H}_2^+,\tilde{W}^-,\tilde{H}_1^-)$, is
\be
M_{{\chi}^{\pm}} =
\left(
\begin{array}{cc}
0&X^t\\
X&0
\end{array}
\right),
\ee
where
\be
X = 
\left(
\begin{array}{cc}
M_2&\sqrt{2}s_{\beta}M_W\\
\sqrt{2}c_{\beta}M_W&-\lambda v_s
\end{array}
\right).
\ee

  For the neutralinos, the mass matrix in basis 
$\psi_i^0=(\tilde{B}^0,\tilde{W}^0,\tilde{H}_1^0,
\tilde{H}_2^0,\tilde{S}$) reads
\be
\mathcal{M}_{\tilde{N}} = 
\left(
\begin{array}{ccccc}
M_1&\cdot&\cdot&\cdot&\cdot\\
0&M_2&\cdot&\cdot&\cdot\\
-c_{\beta}s_wM_Z&\phantom{-}c_{\beta}c_wM_Z&
   0&\cdot&\cdot\\
\phantom{-}s_{\beta}s_wM_Z&-s_{\beta}c_wM_Z&
   \lambda v_s&0&\cdot\\
0&0&\lambda v_2&\lambda v_1&0
\end{array}
\right).
\label{mneut}
\ee

  In our analysis we take $M_1 = \frac{\alpha_1}{\alpha_2}M_2
\simeq \frac{1}{2} M_2$, which
corresponds to what would be expected from universality
at the GUT scale.  With an eye towards electroweak 
baryogenesis, we allow the gaugino masses to have 
a common phase: $M_2 = M\,e^{i\phi}$ with $M$ real. 
This phase also has a significant effect on
the mass of the lightest neutralino.
Since flipping the sign of $\lambda$ is equivalent to 
shifting the gaugino phase by $\phi \to \phi + \pi$,
we will consider only the $\lambda > 0$ case.

  To see how the light neutralino state arises, suppose 
$M_1$ and $M_2$ are very large and real so that the 
gaugino states decouple, leaving only the lower
$3\times3$ Higgsino block.  For $v_1 \ll v_2,v_s$, the 
smallest eigenvalue of this matrix is then 
\be
m_{\tilde{N}_1} \simeq 2\lambda v_1v_2v_s/(v_1^2+v_2^2+v_s^2) ,
\label{mnapprox}
\ee
and the corresponding state is predominantly singlino.
More generally, the mass eigenstates are
\be
\tilde{N}_i \;=\; N_{ij}\;\psi_i^0.
{\label{nmix}}
\ee
where $N_{ij}$ is a unitary matrix such that 
$N^*\mathcal{M}_{\tilde{N}}N^{\dagger}$ 
is diagonal with non-negative entries~\cite{Haber:1984rc}.
We label the states in order of increasing mass
so that $\tilde{N}_1$ is the lightest neutralino.

  Measurements made at LEP~II impose stringent constraints
on the chargino/neutralino spectrum.  Since the coupling
of the charginos to gauge bosons is the same as in the MSSM,
the mass of the lightest chargino must satisfy 
$m_{\chi_1^{\pm}} > 104$~GeV~\cite{Abbiendi:2003sc}.
The corresponding requirement for the neutralinos is 
either $(m_{\tilde{N}_1}+m_{\tilde{N}_2})>209$~GeV,
or $\sigma(e^+e^-\to\tilde{N}_1\tilde{N}_2)\lesssim\,10^{-2}$~fb.
Finally, for $m_{\tilde{N}_1}<M_Z/2$, 
we must have $BR(Z\to\tilde{N}_1\tilde{N}_1) 
< 0.8\times 10^{-3}$~\cite{:2000se}.  

  It is possible to satisfy all of these constraints in
the limit of large $\tan\beta$, in which case
$\tilde{N}_1$ is a very light LSP; $\;m_{\tilde{N}_1} \lesssim 15$~GeV
for $\lambda < 1.0$, $\tan\beta>10$, and $M_2\to\infty$.
This state is mostly singlino, and couples only
weakly to the gauge bosons.
However, this limit also leads to an 
unacceptably large neutralino relic density.  As we will
show in Section~\ref{dm}, for heavy sfermions,
the dominant annihilation channel for
$\tilde{N}_1$ is s-channel Z-exchange\footnote{There are also
contributions to the annihilation cross-section from 
s-channel Higgs exchange, but these processes alone are
not strong enough to produce an acceptable neutralino 
relic density unless the neutralino mass is close
to a half of one of the Higgs boson masses.}.  
For such a light, mostly singlino
$\tilde{N}_1$, the $Z\nt_1\nt_1$ coupling 
is too weak for this state to annihilate efficiently.

  We are thus led to consider values of $\tan\beta$ of order unity.
The $\nt_1$ state now has a sizeable Higgsino component
and correspondingly large couplings to the gauge bosons,
so there is a danger of producing too large a contribution
to the Z-width.  The branching ratio of the $Z$ to two $\nt_1$'s
is given by
\be
BR(Z\to\nt_1\nt_1) = \frac{g^2}{4\pi}\frac{(|N_{13}|^2-|N_{14}|^2)^2}
{24\cos_W^2}\frac{M_Z}{\Gamma_Z}\left[1-\left(\frac{2\,m_{\nt_1}}{M_Z}
\right)^2\right]^{3/2}.
\ee
Combining the branching ratio constraint with that for the
relic density, we find $m_{\nt_1} \gtrsim 35$~GeV is needed
if both conditions are to be met.~(See Section~\ref{dm}.)
As this value depends somewhat on parameters in the Higgs
sector, we impose the weaker constraint $m_{\nt_1} >25$~GeV
in our analysis.

  The magnitude of $\lambda$ must be fairly large, 
$\lambda \gtrsim 0.3$, to raise the mass of the lightest 
neutralino above~$25$~GeV.
($|\lambda| \gtrsim 0.5$ for $m_{\nt_1}>35$~GeV.)
On the other hand, if $\lambda$ is too large it encounters
a Landau pole before the GUT scale.  
This is precisely what happens in the recently proposed 
Fat Higgs model~\cite{Harnik:2003rs}, in which the 
Landau pole corresponds to the Higgs compositeness scale.
We would like to maintain the property of perturbative 
unification in the model (in the usual sense), so we
will focus most of our attention on values of $\lambda$
that remain perturbative up to $M_{GUT} \sim 10^{16}$~GeV.
However, with the Fat Higgs model in mind, we will 
also consider larger values of $\lambda$. 

\begin{figure}[htb]
\centerline{
        \includegraphics[width=0.52\textwidth]{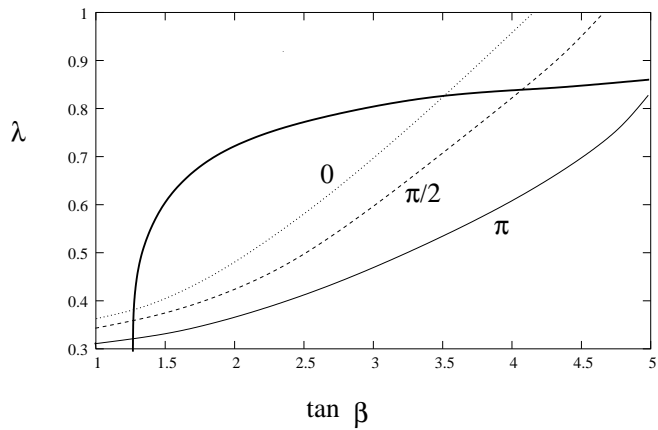}}
        \caption{Allowed regions in the $\tan\beta-\lambda$
plane.  The region consistent with perturbative unification lies
below the thick solid line, while the regions consistent
with the LEP~II constraints (and $m_{\nt_1}>25$~GeV) lie
above the thinner lines.  Among these, the solid line 
corresponds to a gaugino phase of $\phi = \pi$, while
the dotted and dashed lines correspond to $\phi = 0,\:\:\pi/2$
respectively.}
\label{tblam}        
\end{figure}

It is straightforward to derive the limit on $\lambda$
at one-loop order.
The relevant (one-loop) beta functions 
are~\cite{Panagiotakopoulos:2000wp}
\bea
\frac{dg_s}{dt} &=& -\frac{1}{(4\pi)^2}\frac{3}{2}g_s^3,
\nonumber\\
\frac{dy_t}{dt} &=& \frac{y_t}{(4\pi)^2}
(3y_t^2 + \frac{1}{2}\lambda^2 - \frac{8}{3}g_s^2),
\nonumber\\
\frac{d\lambda}{dt} &=& \frac{\lambda}{(4\pi)^2}
(2\lambda^2 + \frac{3}{2}y_t^2), 
\eea
where $t = \ln(Q^2/M_Z^2)$.  Running these up to 
$Q^2 = (10^{16}~\mbox{GeV})^2$ and demanding $\lambda^2 < 4\pi$,
we find the allowed region in the $\tan\beta$-$\lambda$ plane 
shown in figure~\ref{tblam}.  
The lower limit on $\tan\beta$ comes about because 
small values of this quantity imply a large $y_t(m_t)$,
and this accelerates the running of $\lambda$.
The figure also shows
the region in which the $\nt_1$ state has a mass greater than $25$~GeV
and satisfies the LEP~II constraints listed above.

\begin{figure}[htb]
\centerline{
        \includegraphics[width=0.52\textwidth]{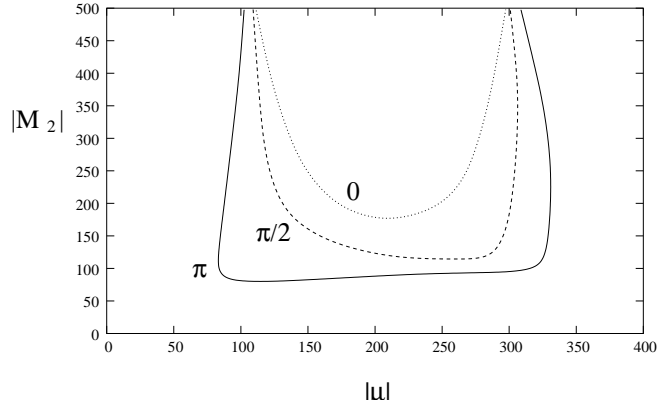}}
        \caption{Allowed regions in the $|\mu|-|M_2|$ plane
for a gaugino phase of $\phi = 0,\:\pi/2,\:\pi$,
and $(\tan\beta,\lambda)$ below the perturbativity bound.
The allowed region lies in the central area.
Recall that $\mu = -\lambda\,v_s$ in the model.}
\label{m2mu}        
\end{figure}

  Figure~\ref{m2mu} shows the corresponding allowed region in the 
$|\mu|\!-\!|M_2|$ plane.  The lower bounds on $|M_2|$ and $|\mu|$ 
are due to the chargino mass constraint. 
Interestingly, there is also an upper bound on $|\mu|$,
which comes from the lower bound on the lightest neutralino mass.
From Eqs.~(\ref{mneut},\ref{mnapprox}) we see that for  
$|v_s| \gg v$, the predominantly singlino state 
becomes very light.
Since $\mu = -\lambda\,v_s$ and $\lambda$
is bounded above by the perturbativity constraint, 
this translates into an upper bound on $|\mu|$.
Both the phase and the magnitude of the gaugino mass $M_2$
have a significant impact on the mass of the light singlino state.
The largest masses are obtained for $\phi = \pi$ with $|M_2| \sim 
\lambda\,v$, as this maximizes the constructive interference
between the gaugino and Higgsino components.  
When $\phi = 0$ the interference is destructive, and $|M_2| 
\to \infty$ maximizes the mass.

\begin{figure}[htb]
\centerline{
\includegraphics[width=0.5\textwidth]{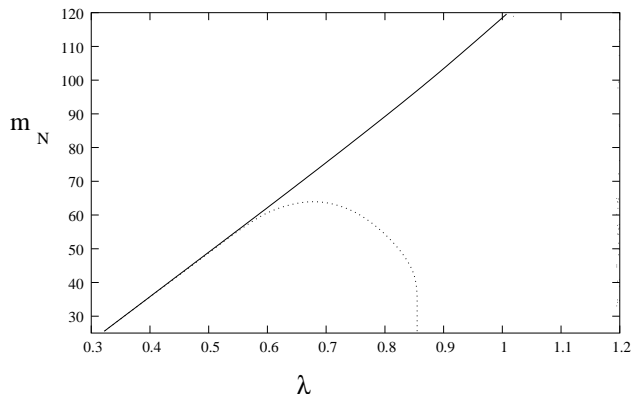}}
\caption{Mass of the lightest neutralino.  The region
to the right of the solid line is consistent with the LEP~II
constraints listed above.  The region surrounded by the dotted
line is also consistent with perturbative unification.}
\label{mn5}        
\end{figure}

  Figure~\ref{mn5} shows the range of masses of the lightest
neutralino that are consistent with the constraints listed above.  
(The relevant input parameter sets are those listed 
in table~\ref{range}).
For $\tan\beta$ and $\lambda$ below the perturbative bound we see
that the $\nt_1$ state has a mass below about $60$~GeV, 
making it the LSP in the absence of a light gravitino.
For $m_{\nt}$ below 50~GeV, this state is predominantly singlino,
with a sizeable Higgsino component.  This is because,
with the assumption of gaugino mass universality, the 
constraint on the chargino mass puts a lower bound on $|M_1|$
that excludes a lighter predominantly Bino state.  
However, a mostly Bino LSP is possible if the light singlino 
state has a mass above about $50$~GeV,
although the parameter space in which this can occur,
consistent with perturbative unification, is severely restricted. 
In this event, the LSP and NLSP must be 
very close in mass.  If $\lambda$ is allowed to exceed
the perturbativity bound, the situation is much less
constrained; the parameter space in which the LSP is 
mostly Bino becomes much larger, and a Bino LSP need no
longer be nearly degenerate with the NLSP. 

\subsection{Higgs Spectrum\label{hgs}}
  The LEP~II lower bound on the mass of the lightest
neutral CP-even Higgs boson of about $114$~GeV 
is difficult to evade in the MSSM.
This follows from the fact that, at tree-level, the mass of this
state is bounded by $M_Z$,
\be
m_h^2 \leq M_Z^2\cos^22\beta,
\ee
which implies that large one-loop corrections are needed to 
increase the mass.
The dominant loop contribution comes from the stops.  With 
$\tan\beta \gg 1$, stop masses of order $1$~TeV, and considerable
fine-tuning of the stop mixing parameters, the mass of the lightest
Higgs can be brought up to 
$m_{h} \simeq 130$~GeV~\cite{Okada:1990vk}.

  The corresponding tree-level bound in the nMSSM
is~\cite{Dedes:2000jp}
\be
m_h^2 \leq M_Z^2(\cos^22\beta + \frac{2\lambda^2}{\bar{g}^2}
\sin^22\beta),
\label{mhup}
\ee
which exceeds $100$~GeV for $|\lambda| \sim 0.7$ and
$\tan\beta \sim 2$.  The same bound applies in the 
NMSSM~\cite{Espinosa:1991gr}.
This makes it possible to avoid the 
LEP~II constraint without fine-tuning in the stop sector. 

  In order to discuss some of the constraints 
on the parameter space from the Higgs sector, 
we list here the tree-level Higgs masses.  
We have also included the one-loop mass corrections from the 
top and the stops given in~\cite{Panagiotakopoulos:2000wp} 
in our numerical analysis.  

  Since the tree-level Higgs VEV's are real, 
and neglecting the small CP-violating 
effects associated with the one-loop chargino and neutralino 
contributions~\cite{Lee:2003nt}, 
the Higgs fields can be expanded as
\bea
 H_1 &=& {v_1 + \frac{1}{\sqrt{2}}(\phi_1 + ia_1)\choose 
\phi_1^-},\nonumber\\
H_2 &=& {\phi_2^+\choose 
v_2 + \frac{1}{\sqrt{2}}(\phi_2 + ia_2)},\nonumber\\
S &=& v_s + \frac{1}{\sqrt{2}}(\phi_s + ia_s).
\eea
After electroweak symmetry breaking, the real and imaginary
parts of the singlet mix with those of $H_1^0$ and $H_2^0$
to produce two neutral scalar states in addition to those
of the MSSM.  In all, the physical Higgs states consist of
one charged scalar, two neutral CP-odd scalars, and three neutral
CP-even scalars.  

  For the CP-odd scalars, the combination $G^0 = 
-a_1\,c_{\beta} + a_2\,s_{\beta}$ is absorbed by the $Z^0$ while
the orthogonal linear combination $A^0 = a_1\,s_{\beta} +
a_2\,c_{\beta}$ mixes with $a_s$ to give two physical scalars.
The mass matrix in basis $(A^0,a_s)$ is
\be
M_P^2 = \left(
\begin{array}{cc}
M_a^2&-a_{\lambda}v\\
-a_{\lambda}v&
-\frac{1}{v_s}(t_s + s_{\beta}\,c_{\beta}\,a_{\lambda}v^2  )
\end{array}
\right),
\label{mhiggsp}
\ee
where 
\be
M_a^2 = -\frac{1}{s_{\beta}\,c_{\beta}}(m_{12}^2 
+ a_{\lambda}\,v_s).
\label{ma}
\ee
Note that Eq.~(\ref{mincond}) implies $m_s^2+\lambda^2\,v^2
= -\frac{1}{v_s}(t_s + s_{\beta}\,c_{\beta}\,a_{\lambda}v^2  )$.
Therefore the singlet soft mass, $m_s^2$, sets the mass
scale of a predominantly singlet state.  

  Among the charged Higgs bosons, the combination $G^+ =  
\phi_2^+\,s_{\beta} - {\phi_1^-}^*\,c_{\beta}$ is taken up by 
the $W^+$ leaving behind a single complex charged 
scalar mass eigenstate, $H^+$, of mass
\be
M_{\pm}^2 = M_a^2 + M_W^2 - \lambda^2v^2.
\ee
It may be shown using the minimization conditions, Eq~(\ref{mincond}),
that $M_{\pm}^2>0$ is equivalent to the condition needed to avoid
a charged Higgs VEV given below Eq.~(\ref{hcharge}).

  Finally, the mass matrix elements for the
CP-even Higgs boson states are
\bea
M_{11}^2 &=& s_{\beta}^2\,M_a^2 + c_{\beta}^2\,M_Z^2\nonumber\\
M_{12}^2 &=& -s_{\beta}\,c_{\beta}
(M_a^2+M_Z^2-2\lambda^2\,v^2)\nonumber\\
M_{13}^2 &=& v(s_{\beta}\,a_{\lambda}
+2c_{\beta}\lambda^2\,v_s)\nonumber\\
M_{22}^2 &=& c_{\beta}^2\,M_a^2+s_{\beta}^2\,M_Z^2\nonumber\\
M_{23}^2 &=& v\,(c_{\beta}\,a_{\lambda}
+2s_{\beta}\lambda^2\,v_s)\nonumber\\
M_{33}^2 &=& -\frac{1}{v_s}(t_s + s_{\beta}\,c_{\beta}\,
a_{\lambda}\,v^2  ) = m_s^2 + \lambda^2v^2
\label{mhiggss}
\eea
with the remaining elements related to these by symmetry.
As for the CP-odd case, the mass of a
mostly singlet state is determined by the singlet soft mass.

  Large values of $M_a^2$ help to increase the mass of the  
Higgs states.  This is most easily obtained with $m_{12}^2 < 0$
(see Eq.~(\ref{ma})), which is a sufficient 
condition to guarantee the absence of spontaneous CP 
violation at tree-level.
We also note that the 
MSSM limit of the NMSSM is not possible in this model.
In this limit one takes $|v_s| \gg v$
while holding $\lambda v_s$ fixed, thereby decoupling the
singlet states from the rest of the Higgs spectrum.
As discussed in Section~\ref{cn}, such large values 
of $v_s$ lead to an unacceptably light neutralino state.
On the other hand,
the decoupling limit of the nMSSM discussed 
in~\cite{Panagiotakopoulos:2000wp}, $|t_s| \to \infty$,
is still viable.  Indeed, the upper bound on the lightest
neutral Higgs mass, Eq.~(\ref{mhup}) is saturated in this limit
if $M_a^2 \to \infty$ as well.  

  The precise LEP~II bounds on the Higgs masses depend on 
the couplings of the Higgs bosons to the gauge bosons.  
These couplings 
tend to be weakened somewhat from mixing with the singlet.  
Let $\mathcal{O}^S,~\mathcal{O}^P$
be the orthogonal mixing matrices relating the gauge and 
mass eigenstates:
\be
\begin{array}{cc}
\left(
\begin{array}{c}
S_1\\S_2\\S_3
\end{array}
\right) = 
\mathcal{O}^S
\left(
\begin{array}{c}
\phi_1\\\phi_2\\\phi_s
\end{array}
\right);
\phantom{aaa}&
\left(
\begin{array}{c}
P_1\\P_2
\end{array}
\right) = 
\mathcal{O}^P
\left(
\begin{array}{c}
A^0\\a_s
\end{array}
\right).
\end{array}
{\label{hmix}}
\ee
We label the mass eigenstates in order of increasing mass,
so that $S_1$ is the lightest CP-even state and $P_1$ the lightest
CP-odd state.
In terms of these mixing matrices, 
the $SVV$-type couplings are
\bea
SZZ:&\phantom{aa}&\frac{\bar{g}}{2}M_Z\,(c_{\beta}\,
{\mathcal{O}^S_{k1}}+ s_{\beta}\,{\mathcal{O}^S_{k2}})
\;(Z_{\mu})^2\,S_k,
\nonumber\\
SWW:&\phantom{aaa}&
\frac{g}{2}M_W\,(c_{\beta}\,{\mathcal{O}^S_{k1}} 
+ s_{\beta}\,{\mathcal{O}^S_{k2}})\;(W_{\mu})^2\,S_k.
\label{hvv}
\eea
Also relevant are the $SPZ$-type couplings
\bea
SPZ:&\phantom{aaa}&\frac{\bar{g}}{2}
\left[{\mathcal{O}^P_{\ell 1}}
(s_{\beta}\,{\mathcal{O}^S_{k1}} - c_{\beta}\,{\mathcal{O}^S_{k2}})
\right]\,Z^{\mu}\:S_k\!\stackrel{\leftrightarrow}{\del}_{\mu}\!
P_{\ell}.
\label{haz}
\eea
The couplings of the Higgs states to fermions and neutralinos
are listed in~Appendix~\ref{appc}.

  The LEP bound on the charged Higgs boson is 
given in Ref.~\cite{:2001xy}.  Assuming $BR(H^+ \to \tau^+ \nu) 
\simeq 1$ this bound reads, approximately,
\be
M_{H^{\pm}} > 90~\mbox{GeV}.
\ee   

The bounds on the CP-odd Higgs bosons depend 
strongly on their coupling to the Z-gauge boson 
and the CP-even scalars given in Eq.~(\ref{haz}). 
If the lightest CP-odd Higgs boson, $P_1$, has a large  
singlet component this coupling can become very
small, and the bound on this particle is much weaker than 
the LEP bound of about 90~GeV present in the MSSM
~\cite{unknown:2001xx}.
This bound may be further weakened if the decay 
$P_1 \to\nt_1\nt_1$ is allowed kinematically.  If so,
this mode tends to dominate the decay width 
leading to a large fraction of invisible final states.

  The same is true of the lightest CP-even state, $S_1$.
The limit found in~\cite{Barate:2003sz}
depends on the strength of the $SVV$ coupling relative
to the corresponding Standard Model coupling.  
From Eq.~(\ref{hvv}) this relative factor
is $|c_{\beta}\,{\mathcal{O}^S_{11}} 
+ s_{\beta}\,{\mathcal{O}^S_{12}}|$, which can
be considerably smaller than unity if the $S_1$ state
has a large singlet component.  Again, the limit
is further weakened if the $S_1\to\nt_1\nt_1$ channel is open,
as this tends to dominate the decay width below 
the gauge boson threshold.
In this case, the limit on invisible decay modes 
found in~\cite{:2001xz} is the relevant one.

\section{Electroweak Baryogenesis\label{ewbg}}
  If Electroweak Baryogenesis~(EWBG) is to generate the
presently observed baryon asymmetry,
the electroweak phase transition must be strongly first order.
In the most promising MSSM scenario, this phase transition
is dominated by a light, mostly right-handed 
stop~\cite{Carena:1996wj}.  Such a stop produces a large 
contribution to the cubic term in the one-loop effective 
potential that is responsible 
for making the phase transition first order.  
Even so, for Higgs masses above the LEP~II bound, 
there is only a very small region of parameter space
in which the EW phase transition is strong enough for 
EWBG to work~\cite{Quiros:2000wk,Carena:2002ss}.  

  The prospects for EWBG in the NMSSM are more promising.
The NMSSM has an additional tree-level contribution to the cubic 
term of the effective potential.  This
is sufficient to make the electroweak phase transition strongly
first order without relying on the contribution of a light stop
~\cite{Pietroni:in,Davies:1996qn,Huber:2000mg,Kang:2004pp}.
Since the nMSSM has a similar cubic term in the tree-level
potential, we expect EWBG to be possible in this model as well.

  In this section we investigate the strength to the electroweak
phase transition in the nMSSM in order to find out whether
a strongly first order transition is possible,
and if so, try to map out the relevant region of parameter space.
To simplify our analysis, we neglect the contributions
from sfermions other than the stops since
these are generally very small.  
We also fix the stop SUSY-breaking
parameters to be
\bea
m_{Q_3}^2 &=& m_{U_3}^2 = (500~\mbox{GeV})^2,\nnmb\\
a_t &=& 100~\mbox{GeV}.\nnmb
\eea
This choice of parameters leads to stops that are too
heavy to have a relevant impact on the strength of the
first order phase transition. 
We have made this choice because it allows us to emphasize the
effects induced by terms in the tree-level effective potential
that are not present in the MSSM. 
These effects turn out to be sufficient to make the phase transition 
strongly first order, even in the absence of light stops. 

  While a strongly first order electroweak phase transition   
is necessary for EWBG to generate the observed baryon asymmetry,
based on previous analyses of the MSSM, 
it appears that this condition is also 
sufficient~\cite{Cline:2000nw,Carena:2000id}. 
In the MSSM, the generation of baryon number 
proceeds from the CP-violating interactions of 
the charginos with the Higgs field.
The dominant source of CP violation, leading to the 
baryon asymmetry, is proportional to the relative phase of the $\mu$
and the gaugino mass parameters, $arg(\mu\,M_2)$. The only difference  
in the model under study is that the $\mu$ parameter is replaced by
the quantity $(-\lambda\,v_s)$, which is real, and 
CP violation is induced by the phase of the gaugino masses. 
Therefore, in the presence of a sufficiently strong first order phase
transition,  we expect a result for the baryon asymmetry 
generated from the chargino currents similar to 
the one obtained in the MSSM.

\subsection{One-Loop Effective Potential\label{oneloop}}
  The finite temperature effective potential for the 
real Higgs scalars is 
\be
 V(\xm_i,T) = V_0(\xm_i) + V_1(\xm_i,T) + V_{daisy}(\xm_i,T)
+ \ldots
\ee
where $V_n$ is the n-loop contribution, and the
additional term, $V_{daisy}$, will be discussed below.
Also, $\xm_i$, $i=1,\,2,\,s$ are the classical field variables 
corresponding to $H_1^0,\,H_2^0,$ and $S$.
($\xm_i = v_i$ at the T=0 minimum.)
The tree-level part is
\bea
V_0(\varphi_i) &=& m_1^2\,\xd^2 + m_2^2\,\xu^2 + m_s^2\,\xs^2 
+ 2\,m_{12}^2\,\xd\xu + 2\,t_s\xs + 2\,a_{\lambda}\,\xs\,\xd\,\xu 
\nonumber\\
&&  
+ \frac{\bar{g}^2}{8}(\xd^2-\xu^2)^2 + \lambda^2\,\xs^2
(\xd^2 + \xu^2) + \lambda^2\,\xd^2\,\xu^2.
\eea
Note the cubic term, $a_{\lambda}\,\xs\,\xd\,\xu$, which has no 
counterpart in the MSSM.  
In the $\overline{DR}$ scheme, the one-loop contribution reads 
\be
V_1(\varphi_i,T) = \sum_{b}g_b\,f_B(m_b^2,T)
  + \sum_{f}g_f\,f_F(m_f^2,T),
\ee
where $b$ runs over bosons, $f$ runs over Weyl fermions, 
and $g_i$ is the number of (on-shell) degrees of freedom.
To a very good approximation,
the functions $f_B,~f_F$ are given by
~\cite{Anderson:1991zb}
\bea
f_B(m^2,T) &=& \left\{
\begin{array}{lc}
-\frac{\pi^2}{90}T^4 + \frac{1}{24}m^2\,T^2 - \frac{1}{12\pi}
(m^2)^{3/2}\,T - \frac{m^4}{64\pi^2}
\ln(\frac{Q^2}{\tilde{a}_BT^2})&; m/T \lesssim 2.2\\
\frac{m^4}{64\pi^2}\left[\ln(\frac{m^2}{Q^2}) - \frac{3}{2}\right]
-(\frac{m}{2\pi T})^{3/2}T^4e^{-m/T}(1 + 
\frac{15}{8}T/m + \ldots)
&; m/T \gtrsim 2.2
\end{array}
\right.
\label{v1b}\nnmb\\
&&\\
f_F(m^2,T) &=& \left\{
\begin{array}{lc}
-\frac{7\pi^2}{720}T^4 + \frac{1}{48}m^2T^2 
+ \frac{m^4}{64\pi^2}\ln(\frac{Q^2}{\tilde{a}_FT^2})
&; m/T \lesssim 1.9\\
-\frac{m^4}{64\pi^2}\left[\ln(\frac{m^2}{Q^2}) - \frac{3}{2}\right]
-(\frac{m}{2\pi T})^{3/2}T^4e^{-m/T}(1 + 
\frac{15}{8}T/m + \ldots)
&; m/T \gtrsim 1.9
\end{array}
\right.
\label{v1f}\nnmb
\eea
Here, $\tilde{a}_B = (4\pi e^{-\gamma_E})^2,~\tilde{a}_F = 
(\pi e^{-\gamma_E})^2$, $m^2$ is the field-dependent
mass at zero temperature.  We neglect $V_2$ and terms
of higher order.

The third contribution to the potential, $V_{daisy}$, 
is a finite-temperature effect~\cite{Kapusta:tk}. 
At zero temperature, one-loop boson self-energy diagrams
which are quadratically divergent in the UV also develop
an IR divergence as the boson mass is taken to zero.
At finite-temperature, when $m\ll T$ the would-be IR
divergence is cut off by $T$ rather than $m$ leading 
to a thermal contribution to the effective mass, $m^2 \to
\overline{m}^2 = m^2 + \alpha\,T^2$.  Resumming the leading
thermal contributions to bosonic self-energies
modifies the effective potential by the amount
\be
V_{daisy} = -\frac{1}{12\pi}\sum_b\;g_b\,\left(\overline{m}_b^2
-m_b^2\right)^{3/2},
\ee
where $\overline{m}^2_b$ is the thermal mass.  
The sum includes gauge bosons,
although only the longitudinal modes of these develop
a thermal contribution to their mass at leading order.
The field-dependent mass matrices relevant
to our analysis (including thermal corrections)
are listed in Appendix~\ref{appa}.

\subsection{Tree-Level Analysis\label{tree}}
  To better understand the effect of the new cubic term,
we have examined a simplified form of the potential  
which allows us to obtain analytic expressions 
for the critical temperature, $T_c$, and the field VEV's.
If the cubic term plays a dominant role in making the electroweak
phase transition first order, we expect this analysis to give
a good qualitative description of the transition. 

  Our first simplifying assumption is that the ratio of the field
values at the broken phase minimum
remains constant up to $T_c$.  That is, we fix $\tan\beta$,
and consider variations in $\xm = \sqrt{\xd^2+\xu^2}$.
To make the one-loop part of the potential more 
manageable, we keep only the leading $\xm^2T^2$ terms in 
the low-temperature expansion, and include only the contributions
of the gauge bosons and the top.  Since the stops are 
assumed to be heavy, the leading temperature-dependent 
contribution comes from the top provided $\lambda$ lies
below the perturbative bound, $\lambda \lesssim 0.8$. 
For larger values of $\lambda$ the contributions from the
Higgs, charginos, and neutralinos become important.  
We shall restrict ourselves to the perturbative regime in 
the present analysis.

In terms of $\xm,~\xs$, and $\beta$, the effective
potential becomes $V = V_0 + V_1$.  The tree-level part is
\be
V_0 = M^2\xm^2 + m_s^2\xs^2 + 2t_s\xs + 2\tilde{a}\xm^2\xs
+ \lambda^2\xm^2\xs^2 + \tilde{\lambda}^2\xm^4,
\ee
where we have defined
\bea
M^2 &=& m_1^2\cos^2\beta + m_2^2\sin^2\beta 
+ m_{12}^2\sin2\beta,\nonumber\\
\tilde{a} &=& a_\lambda\sin\beta\,\cos\beta,\nonumber\\
\tilde{\lambda}^2 &=& \frac{\lambda^2}{4}\sin^22\beta
+ \frac{\bar{g}^2}{8}\cos^22\beta.
\label{trpm}
\eea
Within our approximation, the one-loop part is
\bea
V_1 
&=& \frac{1}{8}\left(g^2 + \frac{\bar{g}^2}{2} 
+ 2y_t^2\sin^2\beta\right)\xm^2T^2.
\eea

To find the minimization conditions, we shall 
make use of the simple (quadratic) $\xs$ dependence 
and  consider only the field-space trajectory 
$\frac{\del V}{\del \xs} = 0$ 
along which the minimum of the potential is found.
This condition allows us to eliminate $\xs$ in terms of $\xm$
giving
\be
 \xs = -\left(\frac{t_s + \tilde{a}\xm^2}
{m_s^2 + \lambda^2\xm^2}\right).
\ee
Inserting this back into the effective potential, we find
\be
V(\xm,T) = m^2(T)\xm^2 - \frac{(t_s+\tilde{a}\xm^2)^2}
{m_s^2 + \lambda^2\xm^2} + \tilde{\lambda}^2\xm^4,
\label{vapprox}
\ee
where $m^2(T) = M^2 + \frac{1}{8}\left(g^2 + \frac{\bar{g}^2}{2} 
+ 2y_t^2\sin^2\beta\right)$.  

  The critical temperature, $T_c\,$, and the VEV at $T_c\,$,
$\:\varphi_c\,$, are
defined by the two conditions
\bea
V(\xm_c,T_c) &=& V(\xm = 0,T_c),\\
\left.\frac{\del V}{\del \xm}\right|_{\xm = \xm_c}&=&0.\nnmb
\eea
Solving for $\xm_c$ and $T_c$ we find
\bea
\xm_c^2 &=& \frac{1}{\lambda^2}
\left(-m_s^2 + \frac{1}{\tilde{\lambda}}
|m_s\,\tilde{a}-\frac{\lambda^2\,t_s}{m_s}|\right),
\nonumber\\
{}\\
T_c^2 &=& 8\left(F(\xm_c^2) - F(v^2)\right)\Big/\left(g^2 
+ \frac{\bar{g}^2}{2} + 2y_t^2\sin^2\beta\right),\nonumber
\label{tc}
\eea 
where 
\be
F(\xm^2) = 2\tilde{a}\left(\frac{t_s+\tilde{a}\xm^2}
{m_s^2 + \lambda^2\xm^2}\right) - \lambda^2
\left(\frac{t_s+\tilde{a}\xm^2}
{m_s^2 + \lambda^2\xm^2}\right)^2 - 2\tilde{\lambda}^2\xm^2.\nnmb
\ee
Both $\xm_c^2$ and $T_c^2$ must of course be positive if a 
solution is to exist.  For $\xm_c^2>0$, we need
\be
m_s^2 < \frac{1}{\tilde{\lambda}}
\left|\frac{\lambda^2\,t_s}{m_s} - m_s\,\tilde{a}\right|.
\label{phic}
\ee
The positivity of $T_c^2$ requires that $F(\xm_c^2)>F(v^2)$.
Since increasing the temperature tends to decrease the field VEV,
this condition will be satisfied if $F(\xm^2)$ is a 
decreasing function
which is the case provided
\be
(m_s\,\tilde{a}-\frac{\lambda^2t_s}{m_s})^2 < \tilde{\lambda}^2(m_s^2+
\lambda^2\xm^2)^3.
\ee
It is sufficient to demand that this hold for 
$\xm = \xm_c$, which gives
\be
m_s^3(m_s^2\tilde{a}-\lambda^2t_s)^2 < \frac{1}{\tilde{\lambda}}
|m_s^2\tilde{a}-\lambda^2t_s|^3.
\ee
This is equivalent to the inequality in Eq.~(\ref{phic}).
Thus, Eq.~(\ref{phic}) is the necessary condition for a 
first order phase transition.  
To satisfy this equation, $m_s^2$
must not be too large.  This can be also seen from 
Eq.~(\ref{vapprox}), which has only positive quadratic and quartic 
terms in the limit $\lambda^2\xm^2 \ll m_s^2$.

\subsection{Numerical Analysis}
  The results of the previous section have been examined 
more carefully by means of a numerical investigation of 
the one-loop effective potential. 
In this analysis, we consider only the dominant 
contributions which are those of the top, the stops, 
the gauge bosons, the Higgs bosons, the charginos, 
and the neutralinos.  
The corresponding field-dependent mass matrices, both at 
zero and finite temperature, are listed in Appendix~\ref{appa}.   
For the purpose of calculating thermal masses, we assume that
the remaining sfermions and the gluino are heavy 
enough to be neglected.  We find that a strongly first order
electroweak phase transition is possible within the nMSSM.

  The procedure we use goes as follows.
To begin, we specify the values of \\
$(\beta,\:v_s,\:a_{\lambda},\:t_s,\:M_a,\:\lambda,\:|M_2|,\:\phi)$,
where $M_2 = |M_2|e^{i\phi}$ is the complex Wino soft mass.
These are chosen randomly from the initial ranges listed
in table~\ref{range}. 
As above, the Bino mass is taken to be $M_1 = M_2/2$,
and we fix the soft stop parameters to be $m_{Q_3}^2 = 
m_{U_3}^2 = (500~\mbox{GeV})^2$, $a_t = 100~$GeV.
The subtraction scale is set at $Q^2 = (150~\mbox{GeV})^2$.
For each parameter set we calculate the mass spectrum 
at zero temperature and impose the experimental 
constraints described in Section~\ref{tz}.
At this point we do not impose any dark matter constraint
other than the necessary condition $m_{\nt_1} > 25$~GeV.
Since, for some parameter sets, the one-loop correction
can destabilize the potential in the $\xs$ direction,
we also check that the minimum at $(\xd,\xu,\xs) = (v_1,v_2,v_s)$ 
is a global minimum at $T=0$.  
Finally, we calculate $\xm_c$ and $T_c$ using the full potential,
where $T_c$ is taken to be the temperature at which 
the symmetric and broken phase minima are equal in depth,
and $\xm_c = \sqrt{\xd^2+\xu^2}$ is the broken phase VEV at this 
temperature.

\begin{table}[hbt!]
\begin{tabular}{|c||c|c|c|c|c|c|c|c|}
\hline
&$\tan\beta$&$\lambda$&$v_s$&$a_{\lambda}$
&$t_s^{1/3}$&$M_a$&$|M_2|$&$\phi$\\
&&&(GeV)&(GeV)&(GeV)&(GeV)&(GeV)&\\
\hline
Range&$1\!\!-\!\!5$&$0.3\!\!-\!\!2.0$&$-\!750\!-\!0$&$0\!-\!1000$
&$0\!-\!1000$&200-1000&$0\!-\!1000$&$0\!-\!\pi$\\
\hline
\end{tabular}
\caption{Ranges of input parameters.\label{range}}
\end{table}

\begin{figure}[hbt!]
\begin{minipage}[t]{0.47\textwidth}
        \includegraphics[width = \textwidth]{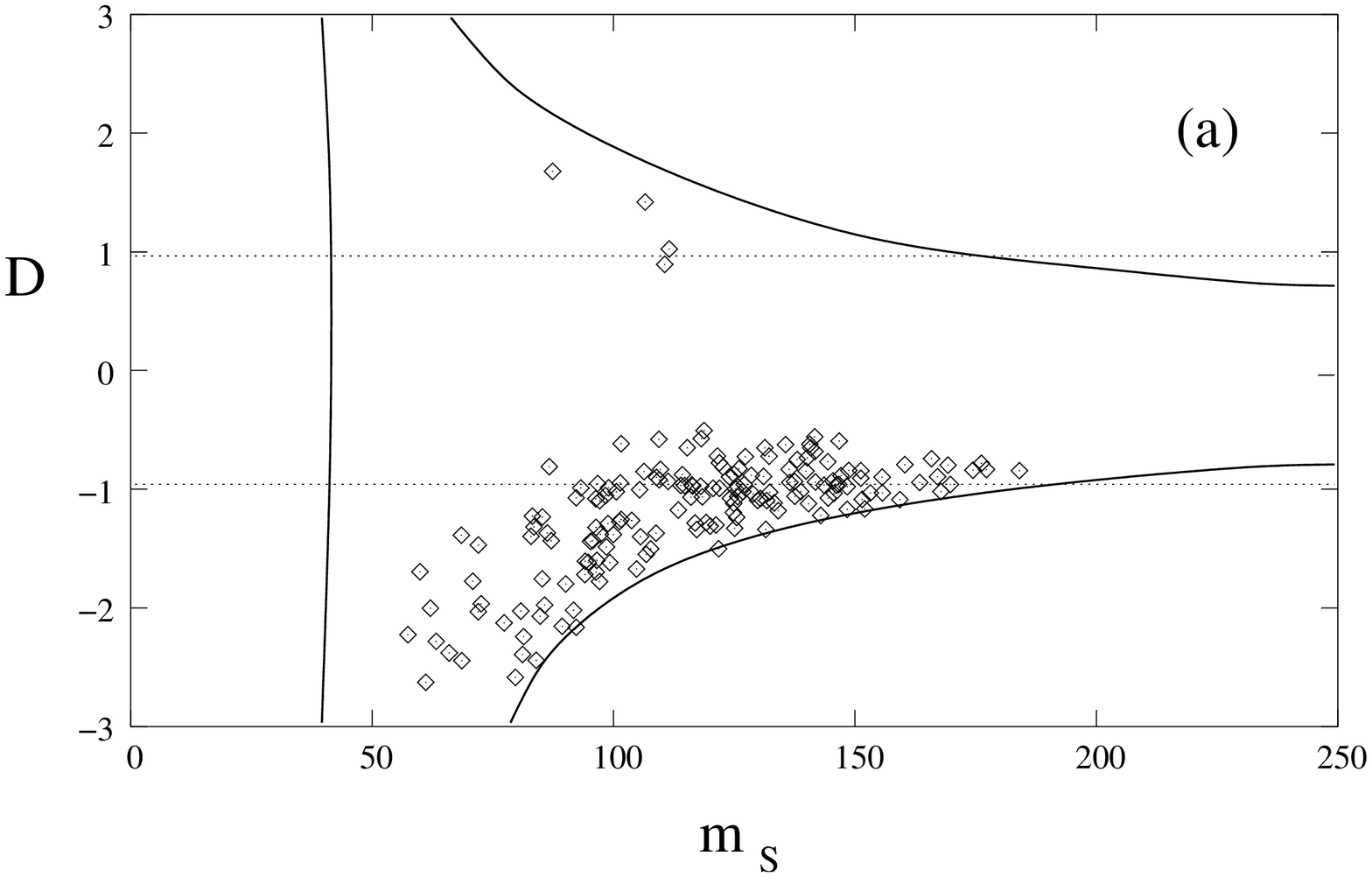}
\end{minipage}
\phantom{aa}
\begin{minipage}[t]{0.47\textwidth}
        \includegraphics[width = \textwidth]{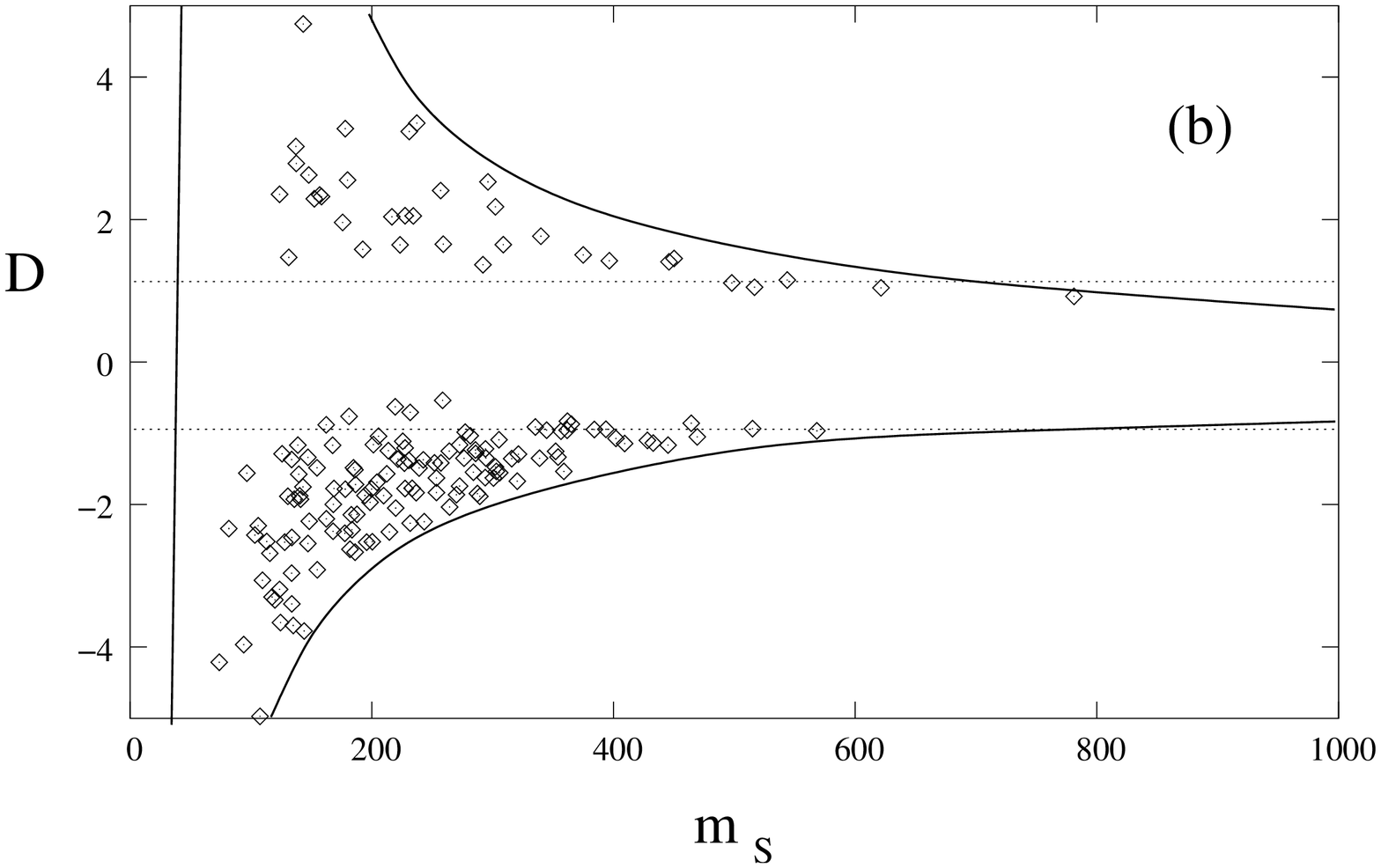}
\end{minipage}
\caption{Values of $D$ for parameter sets leading to a strongly
first order phase transition for:~(a) $\lambda$ below
the perturbative bound;~(b) general values of $\lambda$ in
the range $0.7<\lambda<2.0$.  The region consistent with the
experimental constraints lies within the area enclosed by 
the solid lines.}
\label{pt}
\end{figure}

  In this way we have found several parameter sets which satisfy
all the constraints listed above, and give  
$\xm_c/T_c > 0.9$, which we
take as our criterion for a strongly first order transition
\cite{Carena:1996wj,Cohen:1993nk}.
\footnote{This corresponds to $\frac{\xm_c}{T_c} > 1.3$ for
$\xm$ normalized to $246$~GeV.}  
Let us define 
\be
D = \frac{1}{\tilde{\lambda}\,m_s^2}\left(\frac{\lambda^2\,t_s}{m_s} 
- \tilde{a}\,m_s\right)
\label{dcond}
\ee
where $m_s^2 = -a_{\lambda}\,v_1\,v_2/v_s -t_s/v_s -\lambda^2\,v^2$ 
(see Eq.~(\ref{mincond})),
and $\tilde{a}$ and $\tilde{\lambda}$ are defined
in Eq.~(\ref{trpm}).  $D$ is crucial in 
determining whether or not a first order 
phase transition occurs.
The simplified analysis of Section~\ref{tree}, 
Eq.~(\ref{phic}) in particular, suggests that $|D|>1$  
is a necessary condition for a first order transition.
Figure~\ref{pt} shows $D$ plotted 
against $m_s$ for both $\lambda$ below the perturbative bound,
and for general values in the range $0.7<\lambda<2.0$.
The region surrounded by the solid lines in this figure
corresponds to parameter sets consistent with the 
experimental constraints. 
This figure shows that among the parameter sets
for which a strongly first order phase transition occurs, 
most satisfy $|D|>1$.  
On the other hand, we also find parameter sets with $|D|>1$ 
that do not exhibit a strongly first order phase transition,
so this condition is not a sufficient one.
The low $m_s$ region in these plots is
excluded since the potential tends to become unstable 
in the singlet direction for small values of this quantity.
This leads to the additional requirement of $m_s^2 \gtrsim 
(50~\mbox{GeV})^2$.

  The critical temperature for the phase transition generally
falls in the range $T_c = 100\!-\!150$~GeV.  Table~\ref{sample}
shows the parameter values and transition temperatures for
three of the successful parameter sets.  The particle spectra
corresponding to these are listed in Appendix~\ref{appc}.
Parameter sets $A$ and $B$, with $|D|\simeq~1$, both satisfy 
the perturbative bound while $C$, for which $D \simeq 6.7$,
exceeds it.  

  Sets $A$ and $B$ are typical of the (perturbative)
parameter sets that give a strong phase transition.  
As we found in Section~\ref{cn}, the constraints in the 
chargino/neutralino sector,
along with perturbative consistency,
force $\lambda \sim 0.5\!-0.8$, $\tan\beta \sim 1.5-5$,
and $|v_s| \sim 150\!-\!500$~GeV.  For a given $v_s$, the 
values of $a_{\lambda}$ and $t_s$ must then be adjusted
so that $m_s^2 \gtrsim (50~\mbox{GeV})^2$ (Eq.(\ref{mincond})) and 
$|D| \gtrsim 1$ (Eq.(\ref{dcond})) if the potential 
is to be stable and the transition is 
to be strongly first order.  These quantities are further
constrained by the Higgs sector.  We find that these
requirements may be satisfied for 
$a_{\lambda} \sim 300\!-\!600$~GeV and 
$t_s \sim (50\!-\!200~\mbox{GeV})^3$.  
An upper bound on the value of $m_s$ is obtained, that is about
200~GeV. This bound arises from the phenomenological
constraints and, most importantly, due to the dependence of the
parameter $D$ on $m_s$, Eq. (\ref{dcond}), from
the condition $|D| \simgt 1$. This bound on $m_s$ has important
consequences for Higgs physics, as we will describe in section~\ref{pheno}.

The value of 
$M_a$, instead, does not appear to have much effect on the phase
transition, but tends to be fairly large, $M_a \gtrsim 400$~GeV,
due to the Higgs mass constraints.
While large values of $M_a$ help to increase the masses of the 
lightest Higgs states, they also tend to make EWBG 
less efficient~\cite{Carena:2002ss,Cline:2000nw,Carena:2000id}.
Even so, EWBG is still able to account for the baryon
asymmetry provided $\tan\beta \lesssim 2$, 
as we tend to find here~\cite{Carena:2002ss}.

\begin{table}[htb]
\begin{tabular}{|c||c|c|c|c|c|c|c|c||c|c||c|}
\hline
Set&$\tan\beta$&$\lambda$&$v_s$&$a_{\lambda}$
&$t_s^{1/3}$&$M_a$&$|M_2|$&$\phi$&
$\xm_c$&$T_c$&$\Omega\,h^2$\\
&&&(GeV)&(GeV)&(GeV)&(GeV)&(GeV)&&(GeV)&(GeV)&\\
\hline
A&1.70&0.619&-384&373&157&923&245&0.14&120&125&0.102\\
\hline
B&1.99&0.676&-220&305&143&914&418&2.57&145&95&0.010\\
\hline
C&1.10&0.920&-276&386&140&514&462&2.38&145&130&0.104\\
\hline
\end{tabular}
\caption{Sample parameter sets exhibiting a strongly first
order electroweak phase transition.
\label{sample}}
\end{table}

\section{Neutralino Dark Matter{\label{dm}}}
   As discussed above, the LSP in this model (for $\lambda$
below the perturbative bound) is always the
lightest neutralino, $\nt_1$, with a mass below about $60$~GeV
and a sizeable singlino component.
This can be dangerous since a light, stable particle with 
very weak gauge couplings may produce a relic density 
much larger than is consistent with the observed cosmology.  
On the other hand, if the $\nt_1$ is able to annihilate 
sufficiently well, this state makes a good dark matter 
candidate~\cite{Olive:1990aj}.  

  For values of $\tan\beta$ and $\lambda$ consistent with
the perturbative limit, the LSP tends to be mostly singlino,
but has a sizeable higgsino component.  Since
$m_{\nt_1} \lesssim 60$~GeV, s-channel $Z^0$ exchange 
is the dominant annihilation mode.  There are also 
contributions from s-channel CP-even and CP-odd Higgs 
boson exchanges generated by the 
the $\lambda\,S\,H_1\cdot H_2$ term
in the superpotential, although these tend to be very small except
near the corresponding mass poles.  Since we have assumed that
all sfermions are heavy, we consider only
these two channels in our analysis.  The relevant couplings
are listed in Appendix~\ref{appb}.  (The matrix element for 
s-channel $Z$-exchange is given in~\cite{{Griest:1988ma}}.)  
Let us stress that coannihilation between the lightest neutralino
and the other chargino and neutralino states is not important, since 
for the parameter sets consistent with EWBG
and the perturbative bound, the NLSP is always at least
$15~\%$ (and almost always more than $25\%$) heavier than the LSP, 
implying that the coannihilation contribution is Boltzmann-suppressed.  

  The LSP mass range of interest lies near the Z-pole making
the annihilation cross-section a rapidly varying function of
the mass.  Since this can cause problems for the non-relativistic 
expansion commonly used to calculate the thermal average of the
cross-section~\cite{Griest:1990kh}, we have instead
followed the methods described
in~\cite{Gondolo:1990dk} to do the thermal average.  
This gives
\be
\left<\sigma\,v\right> = \int_{4M^2}^{\infty}ds\,\sqrt{s-4M^2}\,
W\,K_1({\sqrt{s}}/{T})\Big/16\,M^4\,T\,K_2({M}/{T}),
\label{sigv}
\ee
where $M = m_{\nt_1}$ is the neutralino mass, T is the temperature,
$s$ is the usual Mandelstam parameter, $K_1$ and $K_2$ 
are modified Bessel functions, and the quantity W is defined to be
\be
W = \int\left[\prod_f\frac{d^3p_f}{(2\pi)^3\,2E_f}\right]\;
(2\pi)^4\delta^{(4)}(p_1+p_2-\sum_fp_f)\;|\mathcal{M}|^2,
\ee
where $|\mathcal{M}|^2$ is the squared matrix element
averaged over initial states, and summed over final states.

  To find the relic density we have solved the corresponding
Boltzmann equation using the approximation method described 
in~\cite{Kolb:vq}.  The ratio of mass to freeze out temperature,
$x_f = M/T_f$, is given by the solution to
\be
x_f = \ln\left[\frac{0.038\,(g/g_{*_S}^{1/2})\,M\,M_{Pl}
\left<\sigma\,v\right>(x_f)}{x_f^{1/2}}\right],
\ee
where $g=2$ is the number of degrees of freedom of the neutralino,
$g_{*_S}$ is the total number of relativistic degrees of freedom
at temperature $T_f\;$(we use a simple step approximation for 
both $g_*$ and $g_{*_S}$), 
and $M_{Pl}$ is the Planck mass.
The relic density is then given by
\be
\Omega\,h^2 = \frac{(1.07\times10^9\,\mbox{GeV}^{-1})}{M_{Pl}}
\left(\int_{x_f}^{\infty}dx\,\frac{\left<\sigma\,v\right>(x)}{x^2}
g_{*}^{1/2}\right)^{-1}.
\ee

\begin{figure}[hbt!]
\centerline{
        \includegraphics[width=0.65\textwidth]{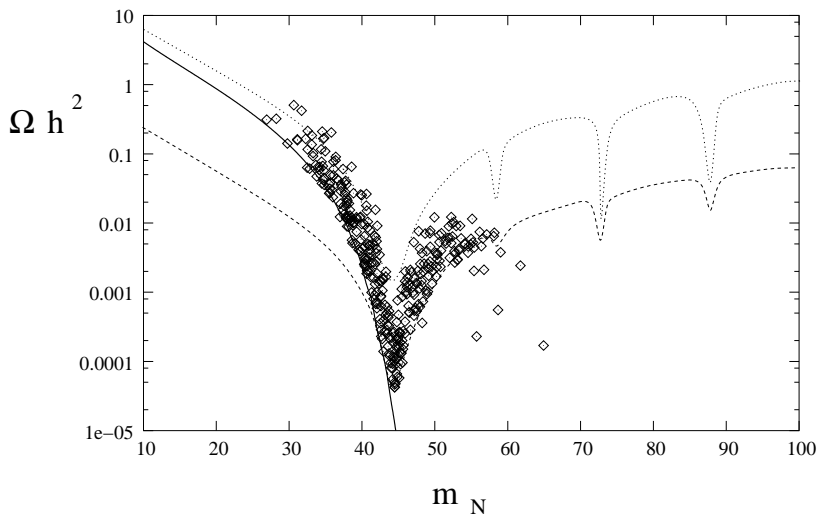}}
        \caption{Neutralino relic density as a function of mass 
for two values of the mixing parameter, 
$\left||{N}_{13}|^2
-|{N}_{14}|^2\right| = 0.1\,\mbox{(dotted)},
\:0.5\,\mbox{(dashed)}$ and typical values of the Higgs 
mixing parameters.  
The region to the right of the thick solid line is 
consistent with the observed $Z$ width.
The scattered points correspond to parameter sets 
that give a strong first order phase transition,
and are consistent with perturbative unification. 
\label{rel2}
}        
\end{figure}
\begin{figure}[hbt!]
\centerline{
        \includegraphics[width=0.65\textwidth]{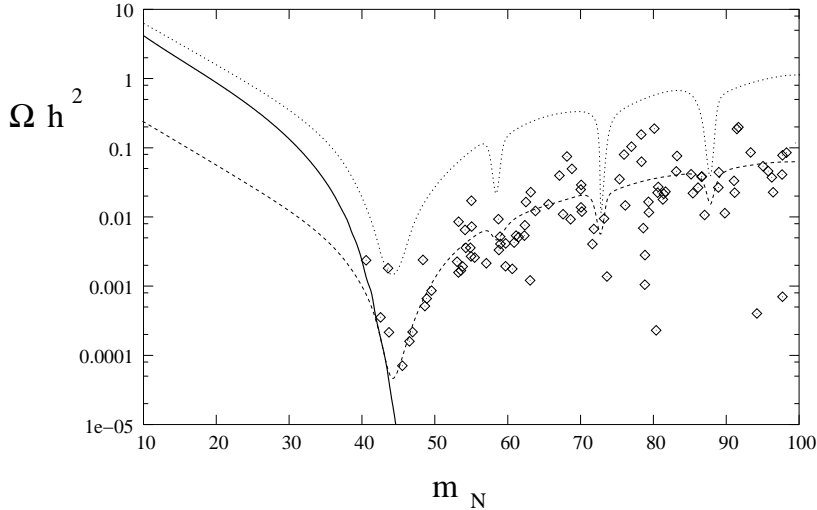}}
        \caption{Neutralino relic density as a function of mass 
for two values of the mixing parameter, 
$\left||{N}_{13}|^2
-|{N}_{14}|^2\right| = 0.1\,\mbox{(dotted)},
\:0.5\,\mbox{(dashed)}$ and typical values of the Higgs 
mixing parameters.  
The region to the right of the thick solid line is 
consistent with the observed $Z$ width.
The scattered points correspond to parameter sets 
which give a strong first order phase transition
with $0.7<\lambda<1.2$. 
\label{rel1}
}                
\end{figure}   

  Figures~\ref{rel2} and \ref{rel1} show the relic 
densities obtained for parameter sets that satisfy 
the abovementioned experimental constraints, and are
consistent with EWBG. 
A relic density 
consistent with the observed dark matter is obtained for 
neutralino masses in the range $m_{\nt_1} \simeq 30\!-\!40$~GeV.
For neutralino masses greater than this, it appears to be
difficult to generate a sufficiently large relic density to
account for the dark matter.
If we allow $\lambda$ to exceed the perturbativity bound,
a realistic dark matter candidate may be obtained for
neutralino masses above about $65$~GeV. 

\section{Phenomenological Discussion\label{pheno}}
  The region of parameter space consistent with EWBG,
neutralino dark matter, and the experimental bounds is
quite constrained, and leads to an interesting phenomenology.  
We shall focus on values of $\tan\beta$ and $\lambda$
that satisfy the perturbativity bound. 

The dark matter condition implies that the LSP of the model
is the lightest neutralino with a mass in the range 
$m_{\nt_1} \simeq 30\!-\!40$~GeV, and is mostly singlino.
For smaller values of $|M_2|$, the next-to-lightest neutralino
is predominantly bino.  Otherwise it is a mostly Higgsino state.
In both cases, there are always two mostly Higgsino states
with masses of order $|\lambda\,v_s|\lesssim 350$~GeV.
The bound on the Higgsino masses comes from the bound
on $|\mu|$ found in Section~\ref{cn}.  This bound also
implies that the lightest chargino state has a mass below
this value.  

  In the Higgs sector, since $M_a$ tends to be fairly large, 
one CP-even Higgs state, one CP-odd Higgs state, and the 
charged Higgs end up with large masses of order $M_a$.
The remaining CP-odd state is relatively light,
and is nearly a pure singlet.  For $M_a \to \infty$, 
the tree-level mass of this state goes to (see~(Eq.~{\ref{mhiggsp}))
\be
m_P \to \sqrt{-(t_s + a_{\lambda}
\,v^2\,s_{\beta}\,c_{\beta})/v_s}= \sqrt{m_s^2 + \lambda^2v^2},
\ee
which is less than $250$~GeV for the values 
of $t_s$, $v_s$ and $a_{\lambda}$ consistent with EWBG,
as may be seen from Figure~\ref{pt}.

  The remaining two CP-even states also tend to be 
fairly light.  In the $M_a \to \infty$ limit, the effective tree-level
mass matrix for these states becomes
\be
M_{S_{1,2}}^2 = \left(
\begin{array}{cc}
M_Z^2\cos^22\beta + \lambda^2\,v^2\,\sin^22\beta&
v(a_{\lambda}\sin2\beta + 2\,\lambda^2v_s)\\
v(a_{\lambda}\sin2\beta + 2\,\lambda^2v_s)&
 -(t_s + a_{\lambda}\,v^2\,s_{\beta}\,c_{\beta})/v_s
\end{array}\right).
\label{mheff}
\ee
Among the parameter ranges consistent with EWBG, 
the off-diagonal element of this matrix can be of the same order 
as the diagonal elements which leads to a strong mixing between
the gauge eigenstates.  However, too much mixing
can produce an unacceptably light mass eigenstate,
in conflict with the LEP bounds~\cite{Barate:2003sz,:2001xz}.
To avoid this, the two terms in the off-diagonal matrix element
must cancel to some extent.  
If $a_{\lambda} = -2\lambda^2\,v_s/\sin2\beta$, the cancellation
is exact, and the mixing goes to zero.  The mass eigenstates 
then consist of the SM-Higgs-like linear combination 
$(\cos\beta\,H_1^0 + \sin\beta\,H_2^0)$ with a tree-level 
mass below $115$~GeV, 
and a singlet state that is degenerate with the lightest 
CP-odd state.  In general these states do mix,
but since the mixing is limited by the Higgs mass constraint,
one state remains predominantly SM-like while the other
is mostly singlet.  For finite $M_a$, the corrections
to this picture are of order $a_{\lambda}v^2/M_a^2$.

   The mass of the SM-like CP-even state is increased significantly
by the one-loop corrections from the top and the stops.
For $m_Q^2 = m_U^2 = 500$~GeV, $a_t = 100$~GeV, as 
used in our analysis, we find that the mass of the lightest
Higgs boson can be as large as 130~GeV, and still be consistent with
EWBG.\footnote{We do not expect two-loop corrections to significantly
alter this result for the set of stop parameters used here.}
This is somewhat larger than the corresponding MSSM limit
of about 120~GeV~\cite{Csaba}, and is the result of the additional
terms in the tree-level potential.  These lead to a larger
Higgs quartic coupling than in the MSSM, 
which sets the scale of the lightest SM-like CP-even state, 
while still allowing for 
a strongly first order electroweak phase transition.  

Figures~\ref{higgsplot} and~\ref{higgscomp} show the 
mass and composition of the three light Higgs states 
(including the one-loop contributions from the top and the stops)
for the representative parameter values 
$M_a = 900$~GeV, $t_s = (150~\mbox{GeV})^3$,
$v_s = -300$~GeV, $a_{\lambda} = 350$~GeV, and $\lambda = 0.7$.
These values are typical of those consistent 
with the constraints and EWBG.
The maximum of the mass of the lightest CP-even state occurs when
the off-diagonal term in Eq.~(\ref{mheff}) vanishes;
$\sin 2\beta = -2\lambda^2\,v_s/a_{\lambda}$. 
The splitting between the $P_1$ and $S_2$ states at this point
is due to the finite value of $M_a$. As $\tan\beta$ varies away
from the maximum point, the mixing between the CP-even Higgs boson
states increases
and, for the specific parameters chosen in Fig.~\ref{higgsplot}, 
values of $\tan\beta \simgt 2.5$ are excluded by the current
Higgs mass bounds obtained by the  LEP 
experiments~\cite{Barate:2003sz,:2001xz}.

\begin{figure}[hbt!]
\centerline{
\includegraphics[width = 0.55\textwidth]{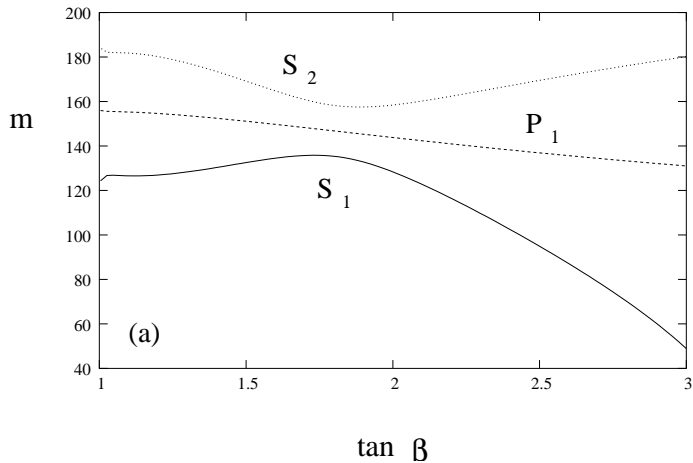}}
\caption{Mass and of the light Higgs bosons
for typical parameter values consistent with EWBG.}
\label{higgsplot}
\end{figure}

\begin{figure}[hbt!]
\centerline{
\includegraphics[width = 0.55\textwidth]{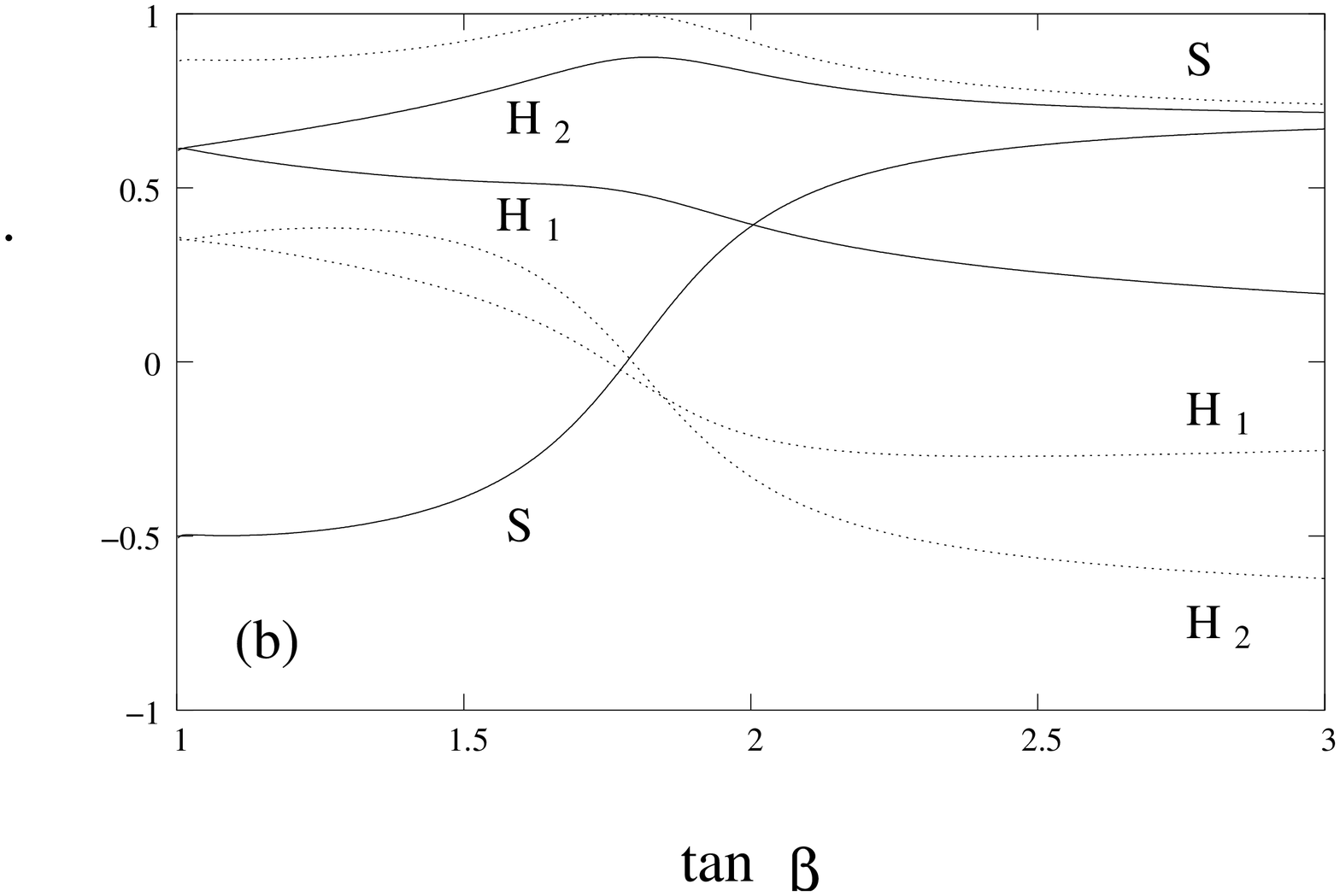}}
\caption{Composition of the two light CP-even Higgs bosons
for typical parameter values consistent with EWBG.
The solid lines correspond to the components of $S_1$,
$\;\mathcal{O}^S_{1i}$, while the dotted lines
are those of $S_2$, $\;\mathcal{O}^S_{2i}$, for $i=1,2,S$.
The composition of the $P_1$ state is not shown because it is
almost pure singlet.}
\label{higgscomp}
\end{figure}

  The discovery of these light states is more challenging than the
MSSM case for two reasons.  First, all three can have sizeable 
singlet components which reduce their couplings to the gauge
bosons and quarks, and therefore their production cross-sections.  
Second, these states can decay invisibly into pairs of the 
neutralino LSP. For CP-even Higgs masses below the weak gauge boson
production threshold this mode dominates the decay 
width:  $BR(S\to\nt_1\nt_1) 
\simeq 0.60-0.95$ for the SM-like state; $BR(S\to\nt_1\nt_1)\simeq 1$
for the mostly singlet state.  However, for masses above (or near) the 
threshold, weak gauge boson final states become
dominant unless the Higgs boson is a nearly pure singlet state.

  Previous studies indicate that the most promising discovery 
mode for an invisibly decaying CP-even Higgs at the LHC is 
vector boson fusion (VBF)~\cite{Choudhury:1993hv,Eboli:2000ze,
Cavalli:2002vs}.  
In~\cite{Cavalli:2002vs} the authors find that, from the
invisible modes alone, Higgs masses up
to $150$~GeV can be excluded at $95\%$~CL with 10~fb$^{-1}$
of integrated luminosity provided $\eta \gtrsim 0.35$, where 
\be
\eta = BR(h\to\mbox{inv})\frac{\sigma(VBF)\phantom{aa}}
{\sigma(VBF)_{SM}}.
\ee
If we treat these limits as being due only to statistical
uncertainty and rescale them by luminosity, we obtain
\be
\mathcal{L}_{95\%} \simeq \frac{1.2~\mbox{fb}^{-1}}{\eta^2},
\hspace{2cm}
\mathcal{L}_{5\sigma} \simeq \frac{8.0~\mbox{fb}^{-1}}{\eta^2},
\ee
where $\mathcal{L}_{95\%}$ and $\mathcal{L}_{5\sigma}$
are the luminosities needed for a 95$\%$~CL exclusion and a $5\sigma$
discovery respectively.
For the SM-like CP-even state, we find $\eta \simeq 0.5-0.9$
among the parameter sets consistent with EWBG.
This implies that, from the invisible channels alone,
less than about $5~\mbox{fb}^{-1}$ of integrated luminosity 
is needed to exclude this state at $95\%$~CL, 
while $10\!-\!30~\mbox{fb}^{-1}$ is 
sufficient for a $5\sigma$ discovery.  
Similarly, we find $\eta\simeq 0\!-\!0.35$ for the mostly singlet 
CP-even state if the mass lies below the gauge boson threshold.  
Thus, at least $10~\mbox{fb}^{-1}$ is needed for a $95\%$~CL exclusion
and $65~\mbox{fb}^{-1}$ for 
a $5\sigma$ discovery using the invisible modes.  
On the other hand, if this state has a mass above the gauge
boson threshold, the Higgs component is usually large enough that 
gauge boson final states completely dominate the branching ratio.  
Lastly, the light CP-odd state is nearly pure singlet 
and tends to decay invisibly, making it extremely 
difficult to observe.

\section{Conclusions\label{conc}}

The origin of the matter-antimatter asymmetry and the
source of the dark matter are two of the most important
questions at the interface of particle physics and 
cosmology. In this article, we have 
investigated these questions within a next-to-minimal 
supersymmetric extension of the Standard Model.
The nMSSM model, which elegantly solves the $\mu$ problem
by adding a gauge singlet superfield, appears
to be consistent with all current experimental constraints,
does not suffer from the usual domain wall problem of the NMSSM,
and leads to the stability of the proton and a neutralino
LSP over cosmological times. 

  We have shown that a strongly first order electroweak 
phase transition, necessary to preserve the baryon asymmetry 
produced by Electroweak Baryogenesis, may be naturally 
obtained within this model.  The strength of the phase
transition is largely determined by terms in the 
tree-level nMSSM scalar potential.  This differs from the 
MSSM, in which one-loop corrections from a light stop are 
needed to make the transition first order.

  We have also shown that, if perturbative consistency is
required to hold up to the GUT scale, the LSP of the model
(in the absence of a light gravitino) 
is always the lightest neutralino, and has a mass
below about 60~GeV.  This state provides
a viable dark matter candidate for neutralino masses
in the range 30-40~GeV.  Furthermore, we find that it is 
possible to achieve simultaneously both a realistic neutralino relic 
density and a strongly first order phase transition.

In the region of parameters consistent with both requirements,
there are always at least two light CP-even and one light 
CP-odd Higgs bosons.  These can decay invisibly into neutralinos 
providing an interesting modification to the standard 
Higgs physics processes.

~\\
{\bf Acknowledgements}

We would like to thank
Csaba Bal\'asz, Marcela Carena,
G\'eraldine Servant, Tim Tait, and Ishai Ben-Dov for useful
discussions. Work at ANL is supported in part by the 
US DOE, Div.\ of HEP,Contract W-31-109-ENG-38.

\newpage

\appendix
\section{Field-Dependent Masses\label{appa}}
 In this section we collect the field and temperature 
dependent mass matrices for those particles relevant to the analysis 
in Section~\ref{ewbg}. The leading thermal mass corrections 
were calculated following~\cite{Brignole:1993wv} 
for vanishing background field
values; $\xd = \xu = \xs = 0$.  Ignoring the singlet
background is reasonable here since, in the parameter 
space of interest, the singlet VEV is closely related 
to the other Higgs VEV's.  To leading order, only bosons 
receive thermal mass corrections.  These come from 
quadratically divergent (at T=0) loops
containing particles which are 
light compared to the temperature; $m\lesssim 2\pi\,T$ 
for bosons and $m\lesssim \pi\,T$ for fermions.  
For the purpose of calculating thermal masses we 
have taken the Higgs bosons,
Higgsinos, electroweak gauginos, and the 
SM particles to be light, while treating the rest of the particles 
in the spectrum as heavy.  We do not expect that a different 
choice of spectrum would change our phase transition 
results since the first order nature of the transition is determined
by the tree-level potential rather than the cubic one-loop term
in $j_B$.  

\subsection{Gauge Bosons}
  At leading order, only the longitudinal components of 
the gauge bosons receive thermal corrections.  The masses are
\bea
m_W^2 &=& \frac{1}{2}g^2(\xd^2+\xu^2) + \Pi_{W^{\pm}}\\
\mathcal{M}^2_{Z\gamma} &=& \left(
\begin{array}{cc}
\frac{1}{2}g^2(\xd^2+\xu^2) + \Pi_{W^3}&-\frac{1}{2}gg'(\xd^2+\xu^2)\\
-\frac{1}{2}gg'(\xd^2+\xu^2)&\frac{1}{2}{g'}^2(\xd^2+\xu^2) + \Pi_{B}
\end{array}
\right),\nnmb
\eea
where $\Pi_i = 0$ for the transverse modes, and 
\bea
\Pi_{W^{\pm}} &=& \frac{5}{2}g^2T^2\\
\Pi_{W^3} &=& \frac{5}{2}g^2T^2\nnmb\\
\Pi_{B} &=& \frac{13}{6}g'^2T^2\nnmb
\eea
for the longitudinal modes.

\subsection{Tops and Stops}
\bea
m_t^2 &=& y_t^2\xu^2\\
\mathcal{M}_{\tilde{t}}^2 &=&
\left(
\begin{array}{cc}
m_Q^2 + m_t^2 + \frac{1}{4}(g^2-\frac{1}{3}{g'}^2)(\xd^2-\xu^2)
+ \Pi_{\tilde{t}_L} &a_t\xu + \lambda\xs\xd\\
a_t\xu + \lambda\xs\xd
&m_U^2 + m_t^2 +\frac{1}{3}{g'}^2(\xd^2-\xu^2)
+ \Pi_{\tilde{t}_R}\nnmb
\end{array}
\right),\nnmb
\eea
where
\bea
\Pi_{\tilde{t}_L} &=& \frac{1}{3}g_s^2T^2 + \frac{5}{16}g^2T^2
+ \frac{5}{432}g'^2T^2 + \frac{1}{6}y_t^2T^2\\
\Pi_{\tilde{t}_R} &=& \frac{1}{3}g_s^2T^2 + 
+ \frac{5}{27}g'^2T^2 + \frac{1}{3}y_t^2T^2.\nnmb
\eea

\subsection{Higgs Bosons}
The CP-even mass matrix elements are
\bea
{\mathcal{M}_{S_{11}}^2} &=& 
m_1^2+\lambda^2(\xu^2+\xs^2)+\frac{\bar{g}^2}{4}(3\xd^2-\xu^2)
+\Pi_{H_1}
\nnmb\\
{\mathcal{M}_{S_{12}}^2} &=& 
m_{12}^2+2\xd\xu(\lambda^2-\frac{\bar{g}^2}{4})+a_{\lambda}\xs
\nnmb\\
{\mathcal{M}_{S_{13}}^2} &=&
a_{\lambda}\xu+2\lambda^2\xd\xs\\
{\mathcal{M}_{S_{22}}^2} &=&  
m_2^2+\lambda^2(\xd^2+\xs^2)+\frac{\bar{g}^2}{4}(3\xu^2-\xd^2)
+\Pi_{H_2}\nnmb\\
{\mathcal{M}_{S_{23}}^2} &=&
a_{\lambda}\xd+2\lambda^2\xu\xs\nnmb\\
{\mathcal{M}_{S_{33}}^2} &=&m_s^2+\lambda^2(\xd^2+\xu^2)+\Pi_{S}
\nnmb
\eea
where the leading thermal corrections are
\bea
\Pi_{H_1} &=& \frac{1}{8}g'^2T^2 + \frac{3}{8}g^2T^2 
+ \frac{1}{12}\lambda^2T^2 \nnmb\\
\Pi_{H_2} &=& \frac{1}{8}g'^2T^2 + \frac{3}{8}g^2T^2 
+ \frac{1}{12}\lambda^2T^2 + \frac{1}{4}y_t^2 \\
\Pi_{S} &=& \frac{1}{6}\lambda^2T^2.\nnmb
\eea

For the CP-odd states we have
\bea
{\mathcal{M}_{P_{11}}^2} &=& 
m_1^2+\lambda^2(\xu^2+\xs^2)-\frac{\bar{g}^2}{4}(\xu^2-\xd^2)
+\Pi_{H_1}
\nnmb\\
{\mathcal{M}_{P_{12}}^2} &=& 
-m_{12}^2-a_{\lambda}\xs
\nnmb\\
{\mathcal{M}_{P_{13}}^2} &=&-a_{\lambda}\xu\\
{\mathcal{M}_{P_{22}}^2} &=&  
m_2^2+\lambda^2(\xd^2+\xs^2)+\frac{\bar{g}^2}{4}(\xu^2-\xd^2)
+\Pi_{H_2}\nnmb\\
{\mathcal{M}_{P_{23}}^2} &=& -a_{\lambda}\xd\nnmb\\
{\mathcal{M}_{P_{33}}^2} &=&m_s^2+\lambda^2(\xd^2+\xu^2)+\Pi_{S}
\nnmb
\eea
where the thermal corrections are as above.

The charged Higgs mass matrix is 
\bea
{\mathcal{M}_{H^{\pm}_{11}}^2}&=& 
m_1^2+\lambda^2\xs^2-\frac{\bar{g}^2}{4}(\xu^2-\xd^2)
+\frac{g^2}{2}\xu^2+\Pi_{H_1}\nnmb\\
{\mathcal{M}_{H^{\pm}_{12}}^2}&=&
-(\lambda^2 - \frac{g^2}{2})\xd\xu - m_{12}^2 - a_{\lambda}\xs\\
{\mathcal{M}_{H^{\pm}_{22}}^2}&=&
m_2^2 +\lambda^2\xs^2+\frac{\bar{g}^2}{4}(\xu^2-\xd^2)
+\frac{g^2}{2}\xd^2+\Pi_{H_2}.\nnmb
\eea

\subsection{Charginos and Neutralinos}
The chargino mass matrix reads  
\be
M_{{\chi}^{\pm}} =
\left(
\begin{array}{cc}
0&X^t\\
X&0
\end{array}
\right)
\ee
where
\be
X = 
\left(
\begin{array}{cc}
M_2&g\xu\\
g\xd&-\lambda\xs
\end{array}
\right).\nnmb
\ee

  For the neutralinos we have
\be
M_{\tilde{N}} = 
\left(
\begin{array}{ccccc}
M_1&\cdot&\cdot&\cdot&\cdot\\
0&M_2&\cdot&\cdot&\cdot\\
-\frac{g'}{\sqrt{2}}\xd&\phantom{-}\frac{g}{\sqrt{2}}\xd&
   0&\cdot&\cdot\\
\phantom{-}\frac{g'}{\sqrt{2}}\xu&-\frac{g}{\sqrt{2}}\xu&
   \lambda\xs&0&\cdot\\
0&0&\lambda \xu&\lambda \xd&0
\end{array}
\right).
\ee
We have taken $M_1 = M_2/2$ and have allowed $M_2$ to be complex
in our analysis.

\section{Higgs and Neutralino Couplings\label{appb}}

We list here the Higgs and Neutralino couplings relevant
to our analysis.  All fermions are written in terms of 
four-component spinors to facilitate the derivation of the 
Feynman rules.  As above, Eq.~(\ref{nmix}), we define 
the unitary matrix
$N_{ij}$ by
\be
\nt_i = N_{ij}\,\psi^0_i
\nnmb
\ee
where $(\psi^0_i) = (\tilde{B^0},\tilde{W^0},\tilde{H^0_1},
\tilde{H^0_2},\tilde{S})$.  Similarly, as in Eq.~(\ref{hmix})
we define the orthogonal matrices $\mathcal{O}^S$ and 
$\mathcal{O}^P$ by
\be
\begin{array}{cc}
\left(
\begin{array}{c}
S_1\\S_2\\S_3
\end{array}
\right) = 
\mathcal{O}^S
\left(
\begin{array}{c}
\phi_1\\\phi_2\\\phi_s
\end{array}
\right);
\phantom{aaa}&
\left(
\begin{array}{c}
P_1\\P_2
\end{array}
\right) = 
\mathcal{O}^P
\left(
\begin{array}{c}
A^0\\a_s
\end{array}
\right).
\end{array}
\nnmb
\ee 

\subsection{Neutralino Couplings}

ZNN:~\cite{Haber:1984rc}
\be
\lag_{ZNN} = \frac{g}{2\cos\theta_W}Z_{\mu}\overline{\nt}_i\,
\gamma^{\mu}
(\mathcal{O}^N_{ij}P_L-\mathcal{O}^{N^*}_{ij}P_R)\nt_j
\ee
where 
\be
\mathcal{O}^N_{ij} = \frac{1}{2}\left(N_{i4}N_{j4}^*
-N_{i3}N_{j3}^*\right).
\nnmb
\ee

  For the Higgs-Neutralino couplings we consider 
only the contribution from the $\lambda SH_1\!\cdot\!H_2$
term in the superpotential.  
We therefore neglect the contribution due to mixing with the gauginos.\\
SNN:
\be
\lag_{SNN} = -\frac{\lambda}{\sqrt{2}}S_k
\overline{\nt}_i\left(\bar{A}^k_{ij}P_L +{A}^k_{ij}P_R\right)
\nt_j,
\ee
where
\bea
A^k_{ij} &=& \mathcal{O}^S_{k1}\,Q^{45}_{ij} 
+ \mathcal{O}^S_{k2}\,Q^{35}_{ij} + \mathcal{O}^S_{k3}\,Q^{34}_{ij},
\nnmb\\
Q^{ab}_{ij} &=& \frac{1}{2}\left(N^*_{ia}N^*_{jb} 
+ N^*_{ib}N^*_{ja}\right).\nnmb
\eea
$\bar{A}^k_{ij}$ is related to ${A}^k_{ij}$ by the replacement
$N^*_{ij}\to N_{ij}$.

PNN:
\be
\lag_{PNN} = -i\frac{\lambda}{\sqrt{2}}S_k
\overline{\nt}_i\left(\bar{B}^k_{ij}P_L -{B}^k_{ij}
P_R\right)\nt_j,
\ee
where
\bea
B^k_{ij} &=& s_{\beta}\,\mathcal{O}^P_{k1}\,Q^{45}_{ij} 
+ c_{\beta}\,\mathcal{O}^P_{k1}\,Q^{35}_{ij} 
+ \mathcal{O}^P_{k2}\,Q^{34}_{ij},
\nnmb
\eea
As with $A^k_{ij}$, $\bar{B}^k_{ij}$ is related to 
${B}^k_{ij}$ by the replacement $N^*_{ij}\to N_{ij}$. 

  Finally we note that in converting these couplings into
Feynman rules, one must insert an additional factor of two
since the neutralinos are written as 
Majorana spinors~\cite{Haber:1984rc}.

\subsection{Higgs Couplings}

  The relevant couplings of the Higgs scalars to the gauge bosons
are given in Section~\ref{hgs} above.  
We list here the couplings of the 
Higgs to the third generation quarks.\\
$S\bar{f}f$:
\be
\lag_{S\bar{f}f} = -\frac{1}{\sqrt{2}}\,y_b\,\mathcal{O}^S_{k1}\,S_k\,
\bar{b}b - \frac{1}{\sqrt{2}}\,y_t\,\mathcal{O}^S_{k2}\,S_k\,
\bar{t}t. 
\ee
$P\bar{f}f$:
\be
\lag_{P\bar{f}f} = 
\frac{i}{\sqrt{2}}\,y_b\,s_{\beta}\,
\mathcal{O}^P_{k1}\,P_k\,\bar{b}\gamma^5b + 
\frac{i}{\sqrt{2}}\,y_t\,c_{\beta}\,\mathcal{O}^S_{k1}\,P_k\,
\bar{t}\gamma^5t.\nnmb 
\ee
The couplings of the Higgs states to the other matter fermions
follow the same pattern.

\subsection{LSP Lifetime}
 
  Having listed the Higgs-Neutralino couplings, we may now
present an estimate for the lifetime of the LSP in the 
$\mathbb{Z}_5^R$ symmetric scenario.  This symmetry
allows the $d=5$ operator $\hat{S}\hat{S}\hat{H_2}\hat{L}$
in the superpotential, which can lead to the decay of the 
neutralino LSP.  This operator generates a coupling which
allows the neutralino to decay into a neutrino and 
a pair of off-shell neutral Higgs scalars,
or an electron, a neutral Higgs scalar, and a charged Higgs
scalar.  We shall focus on the first mode with intermediate
neutral CP-even states and a mostly singlino LSP.  
This gives a more stringent constraint
than the charged mode, and the analysis with intermediate neutral
CP-odd states is analogous.  We will also 
assume that each of the neutral Higgs bosons subsequently decays 
into $\bar{b}b$.  With these assumptions, we find
\be
\Gamma(\nt_1\to\nu\,\bar{b}b\,\bar{b}b) \sim
\frac{\pi}{2(4\pi^2)^4}\frac{y_b^4\left|N_{15}^*
\mathcal{O}^S_{i3}\mathcal{O}^S_{j2}
\mathcal{O}^S_{1i}\mathcal{O}^S_{1j}\right|^2}
{\Lambda^2\,m_H^8}\left(\frac{m_{\nt}}{5}\right)^{11}
\ee
where $\Lambda$ is the cutoff scale, $m_H$ is the Higgs mass
in the intermediate propagators,
and we have set all final state momenta to $m_{\nt}/5$.
Setting the mixing factor to unity, taking $m_{\nt}/m_H \sim 1$
and $\tan\beta = 2$, and demanding that $\Gamma < H_0$,
we find 
\be
\Lambda^2 \gtrsim \left(\frac{m_{\nt}}{\mbox{GeV}}\right)^3
\,10^{23}~\mbox{GeV}^2
\ee
which translates into $\Lambda \gtrsim 3\times 10^{14}$~GeV
for $m_{\nt} = 100$~GeV.

\section{Sample Mass Spectra\label{appc}}
 
We list here the mass spectra for the sample parameter sets
A, B, and C listed in table~\ref{sample}.
\vspace{0.5cm}

Higgs Scalar Masses:
\begin{center}
\begin{table}[htb]
\begin{tabular}{|c||c|c|c||c|c||c|}
\hline
Set&$S_1$&$S_2$&$S_3$&$P_1$&$P_2$&$H^{\pm}$\\
&(GeV)&(GeV)&(GeV)&(GeV)&(GeV)&(GeV)\\
\hline
A&115&158&925&135&927&922\\
\hline
B&116&182&914&164&917&911\\
\hline
C&121&219&504&115&534&498\\
\hline
\end{tabular}
\end{table}
\end{center}

{
Neutralino and Chargino Masses:
\begin{center}
\begin{table}[htb]
\begin{tabular}{|c||c|c|c|c|c||c|c|}
\hline
Set&$\nt_1$&$\nt_2$&$\nt_3$&$\nt_4$&$\nt_5$&
$\chi^{\pm}_1$&$\chi^{\pm}_2$\\
&(GeV)&(GeV)&(GeV)&(GeV)&(GeV)&(GeV)&(GeV)\\
\hline
A&33.3&107&181&278&324&165&320\\
\hline
B&52.4&168&203&221&432&151&432\\
\hline
C&77.1&228&268&331&474&257&474\\
\hline
\end{tabular}
\end{table}
\end{center}
}

{  For completeness, we also display the composition
of the lightest neutralino, 
$\nt_1$, the lightest CP-even Higgs, $S_1$,
and the lightest CP-odd Higgs, $P_1$.
\begin{center}
\begin{table}[htb]
\begin{tabular}{|c||c|c|c|c|c||c|c|c||c|c|}
\hline
Set&$|N_{11}|$&$|N_{12}|$&$|N_{13}|$&$|N_{14}|$&$|N_{15}|$&
$\mathcal{O}^S_{11}$&$\mathcal{O}^S_{12}$&$\mathcal{O}^S_{13}$
&$\mathcal{O}^P_{11}$&$\mathcal{O}^P_{12}$\\
\hline
A&0.13&0.10&0.11&0.37&0.91&0.46&0.74&-0.50&0.08&0.99\\
\hline
B&0.07&0.07&0.16&0.52&0.84&0.42&0.80&-0.44&0.07&0.99\\
\hline
C&0.01&0.01&0.28&0.33&0.90&0.70&0.53&0.48&0.26&0.97\\
\hline
\end{tabular}
\end{table}
\end{center}
}

\end{document}